\documentclass[english,twocolumn,transaction]{IEEEtran}
\usepackage[T1]{fontenc}
\usepackage{float}
\usepackage{amsmath}
\usepackage{dsfont}
\usepackage{amsthm}
\usepackage{amssymb}
\usepackage{stackrel}
\usepackage{graphicx}
\usepackage{setspace}
\usepackage{url}%%in order to show the web link in the reference
\usepackage{color}

\usepackage{caption}
\usepackage{subfigure}
\usepackage{algorithm}
\usepackage{algorithmic}

\makeatletter

\floatstyle{ruled}
\newfloat{algorithm}{tbp}{loa}
\providecommand{\algorithmname}{Algorithm}
\floatname{algorithm}{\protect\algorithmname}

\IEEEoverridecommandlockouts
\usepackage{ifpdf}
\usepackage{cite}
\ifCLASSOPTIONcompsoc
\usepackage[caption=false,font=normalsize,labelfont=sf,textfont=sf]{subfig}
\else
\usepackage[caption=false,font=footnotesize]{subfig}
\fi
\usepackage{amsmath}
\usepackage{mathtools}
\usepackage{booktabs} %{booktabs}
\usepackage{multirow}
\usepackage{setspace}
\usepackage{bm}
\usepackage{mathrsfs}

\usepackage{bbm}
\@ifundefined{showcaptionsetup}{}{%
 \PassOptionsToPackage{caption=false}{subfig}}
\usepackage{subfig}
\makeatother

\usepackage{babel}

\newtheorem{lemm}{Lemma}
\newtheorem{theo}{Theorem}
\newtheorem{remark}{Remark}

\newcommand*{\QEDA}{\hfill\ensuremath{\blacksquare}} % proof blacksquare

\begin{document}
\captionsetup[figure]{font={small}, name={Fig.}, labelsep=period}
\newcommand{\tabincell}[2]{\begin{tabular}{@{}#1@{}}#2\end{tabular}}

\title{Delay-Constrained Joint Power Control, User Detection and Passive Beamforming in Intelligent Reflecting Surface-Assisted Uplink mmWave System}
\author{Yashuai Cao,~\IEEEmembership{Student Member,~IEEE}, Tiejun Lv, \emph{Senior Member, IEEE},\\
Zhipeng Lin, and Wei Ni, \emph{Senior Member, IEEE}

\thanks{

Y. Cao, T. Lv, and Z. Lin are with the School of Information and Communication Engineering,
Beijing University of Posts and Telecommunications (BUPT), Beijing 100876, China (e-mail: \{yashcao, lvtiejun, linlzp\}@bupt.edu.cn).

W. Ni is with Data61, Commonwealth Scientific and Industrial Research,
Sydney, NSW 2122, Australia (e-mail: wei.ni@data61.csiro.au).

}}

\maketitle
\begin{abstract}
While millimeter-wave (mmWave) communications can enjoy abundant bandwidth resource, their high susceptibility to blockage poses serious challenges to low-latency services. In this paper, a novel intelligent reflecting surface (IRS)-assisted mmWave scheme is proposed to overcome the impact of blockage. The scheme minimizes the user power of a multi-user mmWave system by jointly optimizing the transmit powers of the devices, the multi-user detector at the base station, and the passive beamforming at the IRS, subject to delay requirements. An alternating optimization framework is developed to decompose the joint optimization problem into three subproblems iteratively optimized till convergence. In particular, closed-form expressions are devised for the update of the powers and multi-user detector. The IRS configuration is formulated as a sum-of-inverse minimization (SIMin) fractional programming problem and solved by exploiting the alternating direction method of multipliers (ADMM). The configuration is also interpreted as a latency residual maximization problem, and solved efficiently by designing a new complex circle manifold optimization (CCMO) method. Numerical results corroborate the effectiveness of our scheme in terms of power saving, as compared with a semidefinite relaxation-based alternative.
\end{abstract}


\section{Introduction}
Millimeter-wave (mmWave) communications are perceived as a pivotal technology to provide gigabit data-rate in the fifth-generation (5G) mobile systems \cite{7930469, 8676377, 8767965}. In 5G and beyond, the use of mmWave frequencies allows for diverse service-oriented applications in Big Data and Internet of Things (IoT), supporting unprecedentedly low latency for the services. Apart from connectivity, reliability, capacity and latency determined by vertical-specific services, 5G use cases also expect solutions for large coverage and low power consumption communication devices.

Being integrated with new physical-layer techniques, such as massive multi-input multi-output (MIMO) and ultra-dense deployment, mmWave systems provide an appealing solution to the stringent latency and power consumption requirements of many IoT devices and applications.
Take an uplink mmWave system for an example. A joint Europe/Japan H2020 Project is promoting the use of mmWave for mobile edge computing (MEC) \cite{7980746, di2019resilient, 8915829, 8885934, 7962685} for high-capacity and low-latency task offloading. While mmWave can provide unrivaled data rates, the channel intermittency can adversely increase uplink latency and power consumption because mmWave links are highly susceptible to blockage. Barbarossa \emph{et al.} \cite{7980746} proposed two countermeasures to address the issue of blockage, by adopting the plethora of wireless access points and reserving excessive communication resources. Pietro \emph{et al.} \cite{di2019resilient} extended the two countermeasures to optimize the uplink transmit power of mobile users under latency constraints. However, these solutions require the \emph{a-priori} knowledge of blocking probabilities. Moreover, the excessive reservation of network resources and radio equipments also leads to unpalatable costs, penalizing sustainability \cite{8644519}.

Intelligent reflecting surface (IRS), also known as ``reconfigurable intelligent surface (RIS)", has been recently proposed as a promising solution to mitigating the impact of intermittent mmWave links on uplink transmission. Different from active massive MIMO or large intelligent surface \cite{8319526}, IRS is composed of nearly passive low-cost reflecting elements. Originating from reflectarrays and software-defined metamaterials \cite{DBLP:journals/ejwcn/RenzoDHZAYSAHGR19}, IRS can adaptively change the signal propagation by adjusting phase shifters. Specifically, the phase shifter of each IRS element is controlled by a smart controller to reflect incident signals towards a desired direction \cite{8580675}, thanks to recent breakthroughs of radio frequency micro-electromechanical systems (RF-MEMS) and metamaterials \cite{8811733}.
In contrast, conventional reflectarrays typically fabricated with persistent phase shifts have been widely applied in radar and satellite communications.

IRS-aided wireless transmission has been increasingly drawing attention \cite{8741198, 8811733, wu2019joint, wang2019intelligent, 8746155, 2019arXiv190507920G}. In regards of energy efficiency (EE), the authors of \cite{8741198} jointly optimized the transmit power allocation of the base station (BS) and the passive beamforming of an IRS in an IRS-aided downlink multi-user multi-input single-output (MISO) system, where a trade-off between the EE and the number of reflecting elements was observed. The transmit power control of the BS was studied with recommendations on the IRS deployment \cite{8811733}. The transmit power of the BS was also optimized in a multi-IRS-assisted simultaneous wireless information and power transfer (SWIPT) system \cite{wu2019joint}. Later, IRS was applied to create virtual line-of-sight (LoS) to combat the blockages in downlink mmWave links with the signal-to-interference-plus-noise ratio (SINR) maximized \cite{wang2019intelligent}.
The impact of the phase shift design of IRS on network performance was analyzed in \cite{8746155, 2019arXiv190507920G}. The above works have focused on the power control and transmission design in the downlink. Recently, Bai \emph{et al.} \cite{bai2019latency} applied IRS to the uplink transmissions of MEC. The radio and computation resource allocation were jointly optimized to minimize the offloading latency, under the assumption of equal uplink transmit power among the users.
To date, the potential use of IRS to improve the uplink of mmWave systems has yet to be well studied.

Severe blockage can be encountered in mmWave networks. As a consequence, the cost of energy would increase and the latency performance of services would deteriorate. Considering delay-sensitive services, this paper jointly optimizes the power control and passive beamforming of an IRS-assisted uplink mmWave single-input multiple-output (SIMO) system. Distinct from \cite{7980746, di2019resilient, 7962685}, the supplementary links facilitated by the IRS address not only the blockage of mmWave channels but the predefined upload latency requirements of the users.
Different from existing studies based on an equal power assumption in the uplink \cite{bai2019latency}, the considered problem is a joint optimization of the transmit powers of multiple users, the multi-user detectors at an access point (AP) and the configuration of the phase shifts at an IRS. Our key contributions are listed as follows:
\begin{itemize}
\item A new scenario is studied, where the IRS serves as a passive reflector to provide configurable reflecting paths between the AP and multiple simultaneous users. With the aid of the IRS, the stringent latency requirements of the users are met, the co-channel interference between the users is suppressed, and the transmit powers of the users are minimized.
\item An alternating optimization framework is put forth to decouple the originally intractable problem to three tangible subproblems of uplink power control, multi-user detection, and IRS reconfiguration. Closed-form expressions are derived for the user powers and multi-user detectors. The overall convergence of the framework is analyzed.
\item Two efficient algorithms are developed to reconfigure the IRS for passive beamforming. Specifically, we propose a fraction transform-based alternating direction method of multipliers (ADMM) algorithm to optimize the phase shifts of the IRS in parallel. We also transform the configuration of the IRS into a latency residual maximization problem based on a geometric interpretation of constant modulus constraints on passive beamforming. A new Complex Circle Manifold Optimization (CCMO) method is developed to transform the problem to an unconstrained problem on a Riemannian manifold and solve it efficiently using Riemannian gradient descent.
\end{itemize}

The remainder of this paper is organized as follows. Section \ref{sec:system} outlines the system model and formulates the problem. Section \ref{sec:multi} proposes the alternating optimization framework to solve the problem, and presents the closed-form solutions for uplink power control and multi-user detectors. Section \ref{sec:method} develops the two efficient algorithms for the configuration of the phase shifts at the IRS. Section \ref{sec:sim} provides simulation results, demonstrating the significant gains of the proposed approach over a semidefinite relaxation (SDR)-based alternative. Conclusions are  provided in Section \ref{sec:con}.

\emph{Notations}: Lower- and upper-case boldface indicate vector and matrix, respectively. $\{\cdot\}^{\ast}$, $\{\cdot\}^{\mathsf{T}}$, and $\{\cdot\}^{\mathsf{H}}$ stand for conjugate, transpose, and conjugate transpose, respectively. $\mathsf{diag}\{\cdot\}$ stands for diagonalization. $[\cdot]_{i,j}$ denotes the $(i,j)$-th entry of a matrix. $\jmath=\sqrt{-1}$. $\mathsf{Re}\{\cdot\}$, $\mathsf{Im}\{\cdot\}$ and $\mathsf{arg}\{\cdot\}$ denote the real part, imaginary part and phase of a complex value, respectively. $\otimes$ and $\odot$ denote the Kronecker and Hadamard products, respectively. $\mathbbmss{C}$ and $\mathbbmss{R}$ denote the complex space and real space, respectively. $\mathbbmss{E}\{\cdot\}$ stands for expectation. $\Vert \cdot \Vert$ denotes the $\ell_2$ norm. The superscript $(\cdot)^{(t)}$ indicates the $t$-th iteration step. The channel related parameters are listed in Table \ref{tb:symb}.

\begin{table}[t]
\centering\small
\caption{Main Parameters and Their Meanings}
\begin{tabular}{|l|l|}
\hline
 \textbf{Symbol} &  \textbf{Meaning} \\ \hline
 $\mathbf h_{\text{d},k}$  &  CSI between the AP to the $k$-th user \\ \hline
 $\mathbf h_{\text{r},k}$  &  CSI between the IRS to the $k$-th user \\ \hline
 $\mathbf G$               &  CSI between the AP and IRS \\ \hline
 $\mathbf{\Theta}$         &  Passive beamformer of the IRS \\ \hline
 $\mathbf{u}_k$            &  Noise vector of the $k$-th user at the AP \\ \hline
 $\mathbf f_k$             &  Detection vector of the $k$-th user at the AP\\ \hline
\end{tabular}
\label{tb:symb}
\end{table}

\section{System Model}\label{sec:system}
\subsection{IRS Assisted Uplink MmWave SIMO system}
\begin{figure}[t]
	\centering{}\includegraphics[scale=0.5]{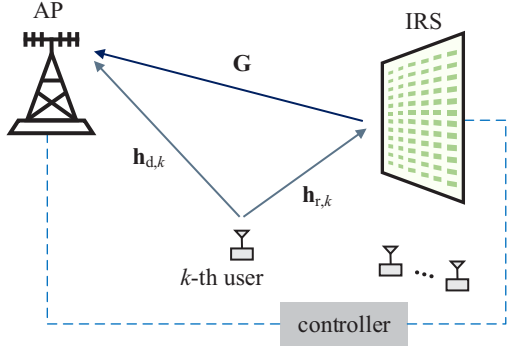}
	\caption{IRS-assisted uplink mmWave system.}
	\label{fig:system}
\end{figure}

Fig. \ref{fig:system} illustrates the proposed IRS-assisted uplink mmWave SIMO system, where $K$ single-antenna users transmit simultaneously to an AP equipped with an $M$-element uniform linear array (ULA). The IRS consists of $N_{\rm az}$ horizontally arranged elements and $N_{\rm el}$ vertically arranged passive reflective elements. $N=N_{\rm az} \times N_{\rm el}$ is the total number of passive elements.
The phase shifters\footnote{The phase shifters are uniformly arranged on a printed circuit board (PCB), and each is associated with embedded tunable chip (e.g., diode or varactor) to tune the impedance or response of the reflecting element \cite{9148961, 8910627, 9122596}.} of the IRS are configured in real-time by a controller.
The AP determines the phase shifts of the IRS, and the results are fed back from the AP to the IRS controller via a dedicated control link.
We assume that the AP has perfect channel state information (CSI) of all relevant mmWave links, as typically assumed in the literature \cite{8741198, 8811733, 8723525, bai2019latency}.

Let $x_{k}= \sqrt{p_k} s_k$ denote the transmit signal from the $k$-th user with $s_k$ and $p_k$ being the unit-power information symbol and the transmit power, respectively. The received signal at the AP from the $k$-th user can be written as
\begin{align}
y_k =& \mathbf f_k^{\sf H} \Big( (\mathbf h_{\text{d},k} + \mathbf G \mathbf{\Theta} \mathbf h_{\text{r},k}  )  {x}_k  \nonumber\\
&+ \sum_{j\neq k}^{K} \left(\mathbf h_{\text{d},j} + \mathbf G \mathbf{\Theta} \mathbf h_{\text{r},j} \right)  {x}_j + \mathbf{u}_k \Big),
\label{eq:signal}
\end{align}
where $\mathbf f_k \in \mathbbmss{C}^{M \times 1}$ is the multi-user detection vector of the $k$-th user and $\mathbf F = [\mathbf f_1, \mathbf f_2, \cdots, \mathbf f_K]$; $\mathbf h_{\text{d},k} \in \mathbbmss{C}^{M \times 1}$ is the CSI between the AP and the $k$-th user; $\mathbf G \in \mathbbmss{C}^{M \times N}$ is the mmWave channel matrix between the AP and IRS; and $\mathbf h_{\text{r},k} \in \mathbbmss{C}^{N \times 1}$ is the CSI between the IRS and the $k$-th user; the phase shift matrix of the IRS is $\mathbf{\Theta}=\mathsf{diag}([\theta_{1}, \cdots, \theta_{N}]^{\sf T})$, where $\theta_{n}=e^{\jmath \varphi_{n}}$ with $\varphi_{n}$ being the $n$-th reflection phase shift; and $\mathbf {u}_k \in \mathbbmss{C}^{M \times 1}$ is the circularly symmetric complex Gaussian (CSCG) noise vector yielding $\mathcal{CN}(0,\sigma_{\text u}^2 \mathbf I)$.

\subsection{Wireless Channel Model}
We assume quasi-static flat-fading mmWave channels.
Based on the widely used 3D Saleh-Valenzuela channel model \cite{6834753, 8269411}, the mmWave channel between the AP and the $k$-th user is given by
\begin{align}
\mathbf h_{\text{d},k}=& \sqrt{ \frac{M}{L+1} } \bigg[ \xi_{k,0} \varrho_{\text{B}}\varrho_{\text{U}} \mathbf a_{M}\left(\phi_{\text{d},k, 0}\right) \nonumber\\
&+ \sum_{\ell=1}^{L} \xi_{k,\ell} \varrho_{\text{B}}\varrho_{\text{U}} \mathbf a_{M}\left(\phi_{\text{d},k, \ell}\right) \bigg],
\label{eq:IRS_user}
\end{align}
where $L$ denotes the number of non-line-of-sight (NLoS) paths; $\ell=0$ indicates the LoS path; $\xi_{k,\ell}$ is the complex channel gain of the $\ell$-th path and $\phi_{\text{d},k, \ell}$ is the associated angle-of-arrival (AoA); $\varrho_{\text{U}}$ and $\varrho_{\text{B}}$ are the transmit and receive antenna element gain, respectively; and $\mathbf a_{M} \in \mathbbmss{C}^{M \times 1}$ is the normalized array steering vector of the ULA at the AP.
We assume that the IRS is properly deployed with high LoS probabilities to both the AP and the users. The paths with two or more reflections are ignored due to severe path loss of mmWave \cite{8811733, 8683145}.
To this end, we assume that the channel between a user and the IRS is LoS, as given by
\begin{equation}
\mathbf h_{\text{r},k}= \sqrt{N} \xi_{k} \varrho_{\text{I}}\varrho_{\text{U}}  \mathbf a_{N}\left(\phi_{\text{r},k}\right),
\label{eq:AP_user}
\end{equation}
where $\xi_{k}$ is the complex channel gain of the $k$-th user and $\mathbf a_{N}\left(\phi_{\text{r},k}\right)$ is the normalized array steering vector between the IRS and the $k$-th user.

The LoS channel between the AP and IRS can be modeled as a rank-one matrix, as given by
\begin{align}
\mathbf G = \sqrt{MN} \xi \varrho_{\text{B}}\varrho_{\text{I}} \mathbf{a}_{M}\left( \phi \right) \mathbf{a}_{N}^{\sf{H}}\left(\vartheta_{\rm{az}}, \vartheta_{\rm{el}}\right), \label{eq:G}
\end{align}
where $\xi$ is the channel gain; $\mathbf{a}_{M}\left( \phi \right) \in \mathbbmss{C}^{M \times 1}$ is the receiver array steering vector at the AP in the direction $\phi$, and $\mathbf{a}_{N}\left(\vartheta_{\rm{az}},\vartheta_{\rm{el}}\right) \in \mathbbmss{C}^{N \times 1}$ is the antenna array steering vector of the transmitter with the elevation angle $\vartheta_{\rm{el}}$ and the azimuth angle $\vartheta_{\rm{az}}$ to the IRS. In (\ref{eq:G}),
\begin{align}
\mathbf{a}_{M}\left( \phi \right) &= \frac{1}{\sqrt{M}} \left[e^{-\jmath \frac{2\pi d}{\lambda} \phi i } \right]_{i \in \mathcal I(M)}, \\
\mathbf{a}_{N}\left(\vartheta_{\rm{az}}, \vartheta_{\rm{el}} \right) &= \mathbf{a}_{N_{\rm az}}\left(\vartheta_{\rm{az}} \right) \otimes \mathbf{a}_{ N_{\rm el}}\left(\vartheta_{\rm{el}} \right),
\end{align}
where $\lambda$ is the wavelength, $d$ is the antenna spacing\footnote{Note that IRS is implemented with discrete antenna elements \cite{8580675, 8644519}, just like uniform rectangular array (URA), where the element spacing of both ULA and IRS is set to be $d = \lambda/2$.}, and $\mathcal I(N_{\delta})=\{n-(N_{\delta}-1)/2, n=0,1,\cdots,N_{\delta}-1\}$.

\subsection{Problem Statement}
The uplink transmission of the $k$-th user is parameterized by $\{D_k, T_k\}$, where $D_k$ is the data size (in nats) and $T_k$ is the maximum latency\footnote{The terminology ``latency'' refers to the over-the-air uplink transmission delay, and the mmWave medium access control (MAC) layer is not discussed here.}. For illustration convenience, we assume that $D_1 \ge D_2 \ge \cdots \ge D_K$ and $T_k = T \ (k=1,2,\cdots, K)$.

A time-division duplexing (TDD) protocol is adopted, and all the CSI can be acquired by channel reciprocity. In this paper, we aim to optimize the total transmit power of the users $\sum_{k=1}^{K}p_k$, the multi-user detector $\mathbf F$, and passive beamformer $\mathbf \Theta$, while satisfying the latency requirements of the users.
The problem is formulated as
\begin{subequations}
\begin{align}
(\text{P1}): \quad  \underset{\mathbf{p}, \mathbf{F}, \mathbf{\Theta}}{\operatorname{min}} \quad & \sum_{k=1}^{K} p_k \nonumber  \\
\text{s.t.}  \quad &   p_k \ge 0, \quad \forall k, \label{eq:MIN1}\\
&  \theta_{n} \in \mathcal F , \quad \ \forall n, \label{eq:MIN2} \\
&   \frac{D_k}{ W \log \left(1 + \Gamma_k \right)} \le T, \ \forall k,\label{eq:MIN3} %quasi-static flat-fadin
\end{align}
\end{subequations}
where $\mathbf p=[p_1,p_2,\cdots, p_K]^{\mathsf T}$ collects the transmit powers of all users, and the continuous feasible set of $\theta_{n}$ is given by
\begin{align}
\mathcal F = \{\theta_{n} = e^{\jmath \varphi_{n}} \vert \varphi_{n} \in [0, 2\pi)\}.
\end{align}
In the latency constraint (\ref{eq:MIN3}), the uplink transmit rate of the $k$-th user is $W \log \left(1 + \Gamma_k \right)$, where $W$ is the channel bandwidth, and the SINR of the $k$-th user is given by
\begin{equation}
\Gamma_k(\mathbf p, \mathbf F, \mathbf \Theta) = \frac{ p_k \left\vert \mathbf f_k^{\sf H} \left(\mathbf h_{\text{d},k} + \mathbf G \mathbf{\Theta} \mathbf h_{\text{r},k} \right) \right\vert^2}{ \sum_{j\neq k}^{K} p_j  \left\vert \mathbf f_k^{\sf H} \left(\mathbf h_{\text{d},j} + \mathbf G \mathbf{\Theta} \mathbf h_{\text{r},j}\right) \right\vert^2 + \sigma_{\text u}^2 \Vert \mathbf f_k \Vert^2 }.
\label{eq:sinr}
\end{equation}

\section{Multi-User mmwave SIMO System}\label{sec:multi}
\subsection{Power Control And Multi-User Detection}
Problem (P1) is intractable due to the nonconvex constant-modulus constraint (\ref{eq:MIN2}) and the coupled optimization variables in constraint (\ref{eq:MIN3}). The problem cannot be solved analytically. To circumvent the impasse, we resort to alternating optimization. Specifically, we alternately optimize each of the variables by fixing the other variables, until convergence.

We start with the uplink transmit power control, given fixed $\mathbf F$ and $\mathbf{\Theta}$.
For notational brevity, we define $\mathbf h_k = \mathbf h_{\text{d},k} + \mathbf G \mathbf{\Theta} \mathbf h_{\text{r},k}$. Thus, (\ref{eq:sinr}) becomes
\begin{equation}
\Gamma_k(\mathbf p) = \frac{ p_k \left\vert \mathbf f_k^{\sf H} \mathbf h_k  \right\vert^2}{ \sum_{j=1,j\neq k}^{K} p_j  \left\vert \mathbf f_k^{\sf H} \mathbf h_j \right\vert^2 + \sigma_{\text u}^2 \Vert \mathbf f_k \Vert^2}.
\label{eq:sinr_p}
\end{equation}
By plugging (\ref{eq:sinr_p}), constraint (\ref{eq:MIN3}) can be rewritten as
\begin{align}
- p_k \left\vert \mathbf f_k^{\sf H} \mathbf h_k  \right\vert^2 + \widetilde{T}_k \left( \sum_{j\neq k}^{K} p_j  \left\vert \mathbf f_k^{\sf H} \mathbf h_j \right\vert^2 + \sigma_{\text u}^2 \Vert \mathbf f_k \Vert^2 \right) \le  0, \ \forall k,  \label{eq:main_cond}
\end{align}
where $\Gamma_k(\mathbf p) \ge \widetilde{T}_k$ with $\widetilde{T}_k = e^{\frac{D_k}{WT}} -1$ being the \emph{minimum protection ratio} of the $k$-th user. (\ref{eq:main_cond}) can be presented in a matrix form, as given by
\begin{align}
\left( \mathbf I- \mathbf Q \right) \mathbf p \succeq \bm \tau,
\end{align}
where
\begin{align}
\left[\mathbf Q\right]_{i,j} &=
\begin{cases}
{0,} & {\text {if}\quad j=i;} \\
{\frac{\widetilde{T}_i \vert \mathbf f_i^{\sf H} \mathbf h_j \vert^2}{\vert \mathbf f_i^{\sf H} \mathbf h_i \vert^2},} & {\text {otherwise},}
\end{cases} \\
\bm \tau &= \left[ \frac{\sigma_{\text u}^2 \widetilde{T}_1 \mathbf f_1^{\sf H}\mathbf f_1}{\vert \mathbf f_1^{\sf H} \mathbf h_1 \vert^2}, \frac{\sigma_{\text u}^2 \widetilde{T}_2 \mathbf f_2^{\sf H}\mathbf f_2}{\vert \mathbf f_2^{\sf H} \mathbf h_2 \vert^2}, \cdots, \frac{\sigma_{\text u}^2 \widetilde{T}_K \mathbf f_K^{\sf H}\mathbf f_K}{\vert \mathbf f_K^{\sf H} \mathbf h_K \vert^2}\right]^{\sf T}.
\end{align}
As a result, problem (P1) can be reformulated as
\begin{subequations}
\begin{align}
(\text{P2}): \quad  \underset{\mathbf{p}}{\min} \quad & \mathbf 1^{\sf T} \mathbf p \nonumber  \\
\text{s.t.} \quad &  \left(\mathbf I- \mathbf Q\right) \mathbf p \succeq \bm \tau, \label{eq:qos}\\
& \mathbf p \succeq \mathbf 0.
\end{align}
\end{subequations}
It is pointed out that $\bm \tau$ in constraint (\ref{eq:qos}) is the vector associated with the \emph{minimum protection ratio} of all users.
We can minimize the objective of (P2) by identifying the critical points of (\ref{eq:qos}). As long as a feasible solution $\mathbf F$ is admitted in the spectral radius of $\mathbf Q$ which is less than unity\footnote{This property can be satisfied, when the specific form of $\mathbf F$ is designed in the following sections.}, the matrix $\mathbf I- \mathbf Q$ is invertible \cite{hiai2014introduction} and $\mathbf p$ can be updated by
\begin{align}
\mathbf p = \left(\mathbf I- \mathbf Q\right)^{-1} \bm \tau. \label{eq:optimal_p}
\end{align}
This property of $\mathbf Q$ provokes a Neumann series expansion \cite{hiai2014introduction}, i.e., $\left(\mathbf I- \mathbf Q\right)^{-1} = \sum_{w=1}^{\infty} \mathbf Q^{w}$. Define the $w$-th iteration as $\mathbf p^{(w)}=\mathbf Q\mathbf p^{(w-1)}+\bm \tau$. The optimal $\mathbf p$ can be obtained recursively with an arbitrary start point $\mathbf p^{(0)}$:
\begin{align}
\underset{w \rightarrow \infty}{\lim}\mathbf p^{(w)} &= \underset{w \rightarrow \infty}{\lim} \bigg\{\mathbf Q^{w}\mathbf p^{(0)} + \bigg[\mathbf Q^{w-1}+\mathbf Q^{w-2}\nonumber\\
&+\cdots+\mathbf Q + \mathbf I\bigg] \bm \tau \bigg\}. \label{eq:series}
\end{align}
In light of (\ref{eq:series}), an equivalent solution can be obtained by taking the limit of the following iterative update process:
\begin{align}
p_k^{(t+1)} = \sum_{j\neq k}^{K} \frac{ \widetilde{T}_k \left\vert \mathbf f_k^{\sf H} \mathbf h_j \right\vert^2}{\left\vert \mathbf f_k^{\sf H} \mathbf h_k  \right\vert^2} p_j^{(t)} +& \frac{\sigma_{\text u}^2 \widetilde{T}_k \Vert \mathbf f_k \Vert^2}{ \left\vert \mathbf f_k^{\sf H} \mathbf h_k  \right\vert^2},\forall k. \label{eq:update_p}
\end{align}
With this iterative scheme, the matrix inversion in (\ref{eq:optimal_p}) can be bypassed for an improved efficiency.

Based on (\ref{eq:update_p}), we can minimize the total transmit power of the users by optimizing $\mathbf{F}$ and $\mathbf{\Theta}$, i.e.,
\begin{align}
\underset{\mathbf{F}, \mathbf{\Theta}}{\arg\min} \sum_{k=1}^{K} \Big\{\sum_{j=1,j\neq k}^{K} & \frac{ \widetilde{T}_k \left\vert \mathbf f_k^{\sf H} \left(\mathbf h_{\text{d},j} + \mathbf G \mathbf{\Theta} \mathbf h_{\text{r},j}\right) \right\vert^2}{\left\vert \mathbf f_k^{\sf H} \left(\mathbf h_{\text{d},k} + \mathbf G \mathbf{\Theta} \mathbf h_{\text{r},k}\right)  \right\vert^2} p_j^{(t)} \nonumber \\
&+ \frac{\sigma_{\text u}^2 \widetilde{T}_k \Vert \mathbf f_k \Vert^2}{ \left\vert \mathbf f_k^{\sf H} \left(\mathbf h_{\text{d},k} + \mathbf G \mathbf{\Theta} \mathbf h_{\text{r},k}\right)  \right\vert^2} \Big\}, \label{eq:BCD}
\end{align}
The block coordinate descent (BCD) method can be employed to optimize $\mathbf{F}$ and $\mathbf{\Theta}$ in an alternating manner.
Given $\mathbf{\Theta}$, the optimization of $\mathbf{F}$ reduces to
\begin{align}
\mathbf f_k &=  \underset{\mathbf f_k}{\arg\min}  \  \frac{\mathbf f_k^{\sf H} \left( \sum_{j=1,j\neq k}^{K} p_j \mathbf D_j +\sigma_{\text u}^2 \mathbf I \right) \mathbf f_k}{\mathbf f_k^{\sf H} \mathbf D_k \mathbf f_k} ,
\  \forall k, \label{eq:MVDR}
\end{align}
where $\mathbf D_j = \mathbf h_j\mathbf h_j^{\sf H}$ for $j=1,2\cdots,K$ are complex Hermitian matrices, and hence the square matrix $\left(\sum_{j=1,j\neq k}^{K} p_j \mathbf D_j+\sigma_{\text u}^2 \mathbf I\right)$ is Hermitian.

We note that (\ref{eq:MVDR}) is in the form of Rayleigh quotient \cite{7151842}, which is related to the generalized eigenvalue problem $\left( \sum_{j\neq k}^{K} p_j D_j + \sigma_{\text u}^2 \mathbf I \right)^{-1} \mathbf h_k \mathbf h_k^{\sf H} \mathbf f_k = \lambda_0 \mathbf f_k \stackrel{C_0 = \mathbf h_k^{\sf H} \mathbf f_k}{\longrightarrow} \frac{C_0}{\lambda_0} \left( \sum_{j\neq k}^{K} p_j D_j+ \sigma_{\text u}^2 \mathbf I \right)^{-1} \mathbf h_k = \mathbf f_k$, where ${\lambda_0}$ is the eigenvalue of $\left( \sum_{j\neq k}^{K} p_j D_j + \sigma_{\text u}^2 \mathbf I \right)^{-1} \mathbf h_k \mathbf h_k^{\sf H}$, and ${C_0}$ is a constant.
The Rayleigh quotient is scale-invariant with respect to $\mathbf{f}_k$, and we can force $C_0=1$.
Thus, the optimization of $\mathbf{f}_k$ is equivalent to the following problem:
\begin{align}
(\text{P3}): \quad  \underset{\mathbf{f}_k}{\min} \quad &  \frac{\mathbf f_k^{\sf H} \left( \sum_{j=1,j\neq k}^{K} p_j \mathbf h_j\mathbf h_j^{\sf H} +\sigma_{\text u}^2 \mathbf I \right) \mathbf f_k}{\mathbf f_k^{\sf H} \mathbf h_k \mathbf h_k^{\sf H} \mathbf f_k} \nonumber\\
\text{s.t.} \quad&   \vert\mathbf f_k^{\sf H} \mathbf h_k \vert= 1, \quad k=1,2,\cdots,K. \label{eq:scale}
\end{align}
The optimal $\mathbf f_k$ in (P3) can be given by the minimum variance distortionless response (MVDR) \cite{725309}, as given by
\begin{align}
\mathbf f_k^{(t+1)}  = \frac{ \left( \sum_{j=1,j\neq k}^{K} p_j^{(t+1)} \mathbf h_j\mathbf h_j^{\sf H} + \sigma_{\text u}^2 \mathbf I \right)^{-1} \mathbf h_k }{ \mathbf h_k^{\sf H} \left( \sum_{j=1,j\neq k}^{K} p_j^{(t+1)} \mathbf h_j\mathbf h_j^{\sf H} + \sigma_{\text u}^2 \mathbf I \right)^{-1} \mathbf h_k }, \label{eq:update_f}
\end{align}
which guarantees that the spectral radius of $\mathbf Q$ is less than one.

\subsection{Problem Transformation of Passive Beamforming}
Given $\mathbf F$, problem (\ref{eq:BCD}) optimizes the reflecting coefficients. We define $\bm \theta =  [\theta_{1}, \cdots, \theta_{N}]^{\sf T}$ and thus $\mathbf\Theta=\mathsf{diag}(\bm\theta)$. Then, we have the following variable substitution:
\begin{align}
\mathbf G \mathbf{\Theta} \mathbf h_{\text{r},j} &= \mathbf G  \cdot \mathsf{diag}\left(\mathbf h_{\text{r},j}\right) \cdot \bm \theta = \mathbf G_{\text{h},j} \bm \theta.
\label{eq:cast_theta}
\end{align}
For notational brevity, we define $b_{k,j} = \mathbf f_k^{\sf H} {\mathbf h}_{\text{d},j}$ and $\mathbf g_{k,j}^{\sf H} = \mathbf f_k^{\sf H}  {\mathbf G}_{\text{h},j}$. Then, (\ref{eq:sinr}) can be rewritten as
\begin{align}
\Gamma_k(\bm \theta)
&= \frac{ p_k \vert b_{k,k} + {\mathbf g}_{k,k}^{\sf H} \bm \theta \vert^2}{ \sum_{j=1,j\neq k}^{K} p_j \vert b_{k,j} + \mathbf g_{k,j}^{\sf H} \bm \theta \vert^2 + \sigma_{\text u}^2 \Vert \mathbf f_k \Vert^2 }.
\label{eq:cast}
\end{align}
By substituting (\ref{eq:cast}) into (\ref{eq:update_p}), the optimization of $\bm{\theta}$, given $\mathbf{F}$ and $\mathbf p$, can be equivalently transformed to
\begin{align}
\bm{\theta}^{(t+1)} =& \underset{ \bm{\theta} \in \mathcal F }{\operatorname{\arg\min}} \sum_{k=1}^{K} \Bigg\{\sum_{j=1,j\neq k}^{K} \frac{ \widetilde{T}_k \vert b_{k,j} + \mathbf g_{k,j}^{\sf H} \bm \theta \vert^2}{\vert b_{k,k} + {\mathbf g}_{k,k}^{\sf H} \bm \theta  \vert^2} p_j  \nonumber\\
&+ \frac{\sigma_{\text u}^2 \widetilde{T}_k \Vert \mathbf f_k \Vert^2}{ \vert b_{k,k} + {\mathbf g}_{k,k}^{\sf H} \bm \theta  \vert^2}\Bigg\}. \label{eq:optimal_theta}
\end{align}
It is arduous to solve (\ref{eq:optimal_theta}) due to the non-convexity of the feasible set of $\bm{\theta}$. We design two efficient algorithms to solve the  subproblem in the next section. The three steps solving $\mathbf p$, $\mathbf F$ and $\mathbf \Theta$ repeat in an alternating manner until a stop condition is met.

\begin{algorithm}[t]
\small
\caption{The proposed alternating optimization framework for uplink power allocation.}
\label{alg:alternating}
\begin{algorithmic}[1]
\REQUIRE {Set feasible values of $\{\mathbf p^{(0)}, \mathbf{F}^{(0)}, \mathbf \Theta^{(0)}\}$ and iteration index $t=0$.}
\REPEAT
\STATE {Set $t \leftarrow t+1$;}
\STATE {With given $\mathbf \Theta^{(t-1)}$, obtain $\mathbf h_k^{(t-1)} = \mathbf h_{\text{d},k} + \mathbf G \mathbf{\Theta}^{(t-1)} \mathbf h_{\text{r},k}, \forall k$;}
\STATE {With given $\mathbf F^{(t-1)}$, $\mathbf p^{(t-1)}$ and $\mathbf h_k^{(t-1)}$, obtain $p_k^{(t)}$ using (\ref{eq:update_p});}
\STATE {Set iteration index $l = 0$, $\mathbf p^{(l)}=\mathbf p^{(t)}$, $\mathbf F^{(l)}=\mathbf F^{(t-1)}$, $\mathbf \Theta^{(l)}=\mathbf \Theta^{(t-1)}$ and $\mathbf h_k^{(l)}=\mathbf h_k^{(t-1)}$;}
\REPEAT
\STATE {With given $\mathbf p^{(l)}$ and $\mathbf h_k^{(l)}$, update $\mathbf f_k^{(l+1)}$ using (\ref{eq:update_f});}
\STATE {With given $\mathbf p^{(l)}$ and $\mathbf f_k^{(l+1)}$, update $\mathbf \Theta^{(l+1)}$ by solving problem (\ref{eq:optimal_theta});}
\STATE {Update $\mathbf h_k^{(l+1)} = \mathbf h_{\text{d},k} + \mathbf G \mathbf{\Theta}^{(l+1)} \mathbf h_{\text{r},k}$;}
\STATE {Set $l \leftarrow l+1$;}
\STATE {With given $\mathbf F^{(l-1)}$, $\mathbf p^{(l-1)}$ and $\mathbf h_k^{(l-1)}$, obtain $p_k^{(l)}$ using (\ref{eq:update_p});}
\UNTIL {$\vert \mathbf 1^{\sf T} \mathbf p^{(l)} - \mathbf 1^{\sf T} \mathbf p^{(l-1)} \vert$ converges.}
\STATE {Obtain $\mathbf p^{(t)}=\mathbf p^{(l)}$, $\mathbf F^{(t)}=\mathbf F^{(l)}$ and $\mathbf{\Theta}^{(t)}=\mathbf{\Theta}^{(l)}$.}
\UNTIL {$\vert \mathbf 1^{\sf T} \mathbf p^{(t)} - \mathbf 1^{\sf T} \mathbf p^{(t-1)} \vert$ converges.}
\STATE {Output optimal $\{\mathbf p, \mathbf{F}, \mathbf \Theta \}$ and calculate $\sum_{k=1}^{K} p_k$.}
\end{algorithmic}
\end{algorithm}

Algorithm \ref{alg:alternating} summarizes the proposed joint optimization of the transmit powers, multi-user detector, and IRS configuration. The convergence analysis of Algorithm 1 follows.
We can write the minimum of (\ref{eq:BCD}) as a function of the optimal $\mathbf F$ and $\bm{\theta}$, as given by
\begin{align}
\mathcal P \left( \mathbf{F}, \bm{\theta} \right) = \underset{\mathbf{F}, \bm{\theta}}{\min} \sum_{k=1}^{K} \Big\{\sum_{j\neq k}^{K} & \frac{ \widetilde{T}_k \left\vert \mathbf f_k^{\sf H} \left(\mathbf h_{\text{d},j} + \mathbf G_{\text{h},j} \bm \theta\right) \right\vert^2}{\left\vert \mathbf f_k^{\sf H} \left(\mathbf h_{\text{d},k} + \mathbf G_{\text{h}, k} \bm \theta\right)  \right\vert^2} p_j  \nonumber \\
&+ \frac{\sigma_{\text u}^2 \widetilde{T}_k \Vert \mathbf f_k \Vert^2}{ \left\vert \mathbf f_k^{\sf H} \left(\mathbf h_{\text{d},k} + \mathbf G_{\text{h},k} \bm \theta\right)  \right\vert^2} \Big\}. \label{eq:map_k}
\end{align}
Define $\mathcal M_{\mathbf{f}_k}( \mathbf{p} ) = \underset{\mathbf f_k}{\min}  \  \sum_{j\neq k}^{K} \frac{ \widetilde{T}_k \vert \mathbf f_k^{\sf H} \mathbf h_j \vert^2}{\vert \mathbf f_k^{\sf H} \mathbf h_k  \vert^2} p_j^{(t)} + \frac{\sigma_{\text u}^2 \widetilde{T}_k \Vert \mathbf f_k \Vert^2}{\vert \mathbf f_k^{\sf H} \mathbf h_k \vert^2}$. Then, $\mathcal P \left( \mathbf{F}, \bm{\theta} \right) = \underset{\bm{\theta}}{\min} \sum_{k=1}^{K} \mathcal M_{\mathbf{f}_k}( \mathbf{p} )$.
Upon the convergence of the BCD iterations, the optimal multi-user detectors keep $\mathcal M_{\mathbf{f}_k}$ non-increasing, while $\mathbf f_k$ is being updated.

\begin{lemm}
The iterative update of the users' transmit powers yields a convergent and unique fixed point.
\end{lemm}
\begin{IEEEproof}
See Appendix \ref{prf1}.
\end{IEEEproof}

\begin{remark}
When the maximum power limit of each user, denoted by $P_{\max}$ is considered, an additional feasibility check step of problem (P1) is required \cite{6884852}. The proposed approach can be potentially extended to capture the power limits. One possible approach is that a penalizing coefficient $w_k^{(t-1)}=\frac{\max \{p_k^{(t-1)}, P_{\max}\}}{P_{\max}}$ can be multiplied to the transmit power of the $k$-th user, $p_k$, in the objective of problem (P1), where $p_k^{(t-1)}$ is the result after running Algorithm 1 for the $(t-1)$-th time. Because the objective is a linear weighted sum of $p_k$, we can run Algorithm 1 to obtain the transmit powers of the users, the multi-user detector at the AP, and the phase shifts of the IRS.
If the resulting transmit powers, i.e., $p_k^{(t)},\forall k$, violate the power limits of the users, the current set of penalizing coefficients, $w_k^{(t-1)}\, \forall k$, make the problem infeasible. We can update the penalizing coefficients and run Algorithm 1 again. This can be done recursively until the problem becomes feasible; or, the problem is infeasible.
The penalizing coefficients magnify the transmit powers of the users which do not meet their power limit requirement in the objective of problem (P1), raise the stake of those users in the minimization of the total (weighted) power, and hence push the transmit powers of those users towards $P_{\max}$. It will be our future work to analyze the efficiency of this approach.
\end{remark}

\section{Reflection Beamforming}\label{sec:method}
\subsection{SIMin Fraction Transform Based ADMM}
Typically, there are a large number of reflecting elements in an IRS. Algorithms suitable for parallel implementations are highly desirable. In this paper, the original passive beamforming problem (\ref{eq:optimal_theta}) is first transformed to a fractional programming problem. Then, we propose a new algorithm that integrates the fraction transform-based alternating optimization and the ADMM and is particularly tailored for the problem.

Given $\mathbf p$ and $\mathbf F$, the optimization of $\mathbf \Theta$ is recast as
\begin{align}
(\text{P4}): \quad  \underset{\bm \theta}{\operatorname{min}} \quad &   \sum_{k=1}^{K} \widetilde{T}_k \frac{ \sum_{j\neq k}^{K} p_j \left\vert b_{k,j} + \mathbf g_{k,j}^{\sf H} \bm \theta \right\vert^2 + \sigma_{\text u}^2 \Vert \mathbf f_k \Vert^2 }{\left\vert b_{k,k} + {\mathbf g}_{k,k}^{\sf H} \bm \theta  \right\vert^2} \nonumber\\
\text{s.t.} \quad&  \text{(\ref{eq:MIN2})}. \nonumber
\end{align}
Problem (P4) is a \emph{weighted sum of inverse SINR minimization} problem. Typically, fractional programming problems can be solved by decoupling numerators and denominators. However, conventional decoupling methods, such as Dinkelbach's algorithms, cannot resolve the sum-of-ratio problems directly \cite{8314727, 6747287, wang2019minimization}. Moreover, the complex variables $\bm \theta$ make the sum-of-inverse minimization (SIMin) problem even more challenging. In this paper, we develop a new fraction transform technique, termed SIMin fraction transform, in the following theorem.

\begin{theo}
\label{theo:fp}
Given $K$ pairs of positive functions $A_k(\bm \theta)$ and $B_k(\bm \theta)$, a sum-of-inverse fractional minimization problem is given by
\begin{align}
\underset{\bm \theta \in \mathcal F}{\operatorname{min}} \quad &  \sum_{k=1}^{K} \frac{A_k(\bm \theta)}{B_k(\bm \theta)},
\end{align}
which is equivalent to
\begin{align}
\underset{\bm \theta \in \mathcal F}{\operatorname{min}} \quad &  \sum_{k=1}^{K} z_k {A_k(\bm \theta)}^2 + \sum_{k=1}^{K} \frac{1}{4 z_k} \frac{1}{B_k(\bm \theta)^2}.
\end{align}
where $\mathbf z = \left[z_1, z_2,\cdots, z_K \right]^{\sf T}$ is an auxiliary vector.
\end{theo}
\begin{IEEEproof}
See Appendix \ref{prf2}.
\end{IEEEproof}

Based on Theorem 1, problem (P4) can be equivalently cast as
\begin{align}
\underset{\bm \theta}{\operatorname{min}} \quad & J(\bm \theta)  = \sum_{k=1}^{K} J_{A,k} (\bm \theta) + \sum_{k=1}^{K} J_{B,k} (\bm \theta), \label{eq:opt_theta}
\end{align}
where
\begin{align}
J_{A,k} (\bm \theta) &= \beta_k  {\widetilde{T}_k^2 \Big( \sum_{j\neq k}^{K} p_j \left\vert b_{k,j} + \mathbf g_{k,j}^{\sf H} \bm \theta \right\vert^2 + \sigma_{\text u}^2 \Vert \mathbf f_k \Vert^2 \Big)^2 }, \\
J_{B,k} (\bm \theta) &= \frac{1}{4\beta_k} \frac{1}{\left\vert b_{k,k} + {\mathbf g}_{k,k}^{\sf H} \bm \theta  \right\vert^4},
\end{align}
where $\bm \beta = \left[\beta_1, \beta_2, \cdots, \beta_K \right]^{\sf T}$ is the auxiliary vector.
According to Theorem \ref{theo:fp} and Appendix \ref{prf2}, the optimal $\beta_k$ is given by
\begin{align}
\beta_k = \frac{1}{2{\widetilde{T}_k {\vert b_{k,k} + {\mathbf g}_{k,k}^{\sf H} \bm \theta  \vert^2} } \Big( \sum_{j\neq k}^{K} p_j \left\vert b_{k,j} + \mathbf g_{k,j}^{\sf H} \bm \theta \right\vert^2 + \sigma_{\text u}^2 \Vert \mathbf f_k \Vert^2 \Big)}.
\label{eq:up_beta}
\end{align}
Next, we decouple the optimization variables between $J_{A,k} (\bm \theta)$ and $J_{B,k} (\bm \theta)$, by employing the ADMM method. In particular, the augmented Lagrangian of (\ref{eq:opt_theta}) is given by
\begin{align}
\mathcal L_{\rho} (\bm \theta, \bm q, \bm r ) =& \sum_{k=1}^{K} J_{A,k} (\bm \theta) + J_{B,k} (\bm q) \nonumber\\
&+ \frac{\rho}{2} \left\Vert \bm \theta- \bm q + \bm r \right\Vert^2  - \sum_{n=1}^{N} \mathbbmss{1}_{\mathcal F}(\theta_n), \label{eq:admm_l}
\end{align}
where $\rho$ is the penalty parameter, and the indicator function of the feasible set of $\bm \theta$ is given by
\begin{align}
\mathbbmss{1}_{\mathcal F}(\theta_n) =
\begin{cases}
0,      & {\theta_n \in \mathcal F,}\\
+\infty, & \text{otherwise}.
\end{cases}
\end{align}
By differentiating (\ref{eq:admm_l}), the ADMM solver for (\ref{eq:opt_theta}) is developed to alternately updating (\ref{eq:up1}), (\ref{eq:up2}) and (\ref{eq:up3}) until convergence.
The sequential iterative form is given by
\begin{align}
\bm{\theta}^{(l+1)} &:= \underset{\theta_n \in \mathcal F}{\arg\min} \sum_{i=1}^{K} J_{A,k} (\bm \theta) + \frac{\rho}{2} \left\Vert \bm \theta- \bm q^{(l)} + \bm r^{(l)} \right\Vert^2, \label{eq:up1} \\
\bm{q}^{(l+1)} &:= \underset{\bm{q}}{\arg\min} \sum_{i=1}^{K} J_{B,k} (\bm q)+\frac{\rho}{2}\left\Vert\bm{\theta}^{(l+1)}-\bm{q}+\bm{r}^{(l)}\right\Vert^{2}, \label{eq:up2}\\
\bm{r}^{(l+1)} &:= \bm{r}^{(l)}+\bm{\theta}^{(l+1)}-\bm{q}^{(l+1)}. \label{eq:up3}
\end{align}

To solve (\ref{eq:up1}), we relax constraint (\ref{eq:MIN2}) as
\begin{align}
\bm \theta^{\sf H} \mathbf e_n \mathbf e_n^{\sf H} \bm \theta \le 1, \quad \forall n=1,2,\cdots, N. \label{eq:cvx_cond}
\end{align}
By replacing (\ref{eq:MIN2}) with (\ref{eq:cvx_cond}), problem (\ref{eq:up1}) can be rewritten as
\begin{align}
\underset{ \bm\theta, \bm\varepsilon}{\min} \quad \mathcal G_1\left(\bm \theta, \bm\varepsilon \right)=&\sum_{k=1}^{K} J_{A,k} (\bm \theta) + \frac{\rho}{2} \left\Vert \bm \theta- \bm q + \bm r \right\Vert^2 \nonumber\\
&+ \sum_{n=1}^{N} \varepsilon_n \left( \bm \theta^{\sf H} \mathbf e_n \mathbf e_n^{\sf H} \bm \theta - 1 \right),
\end{align}
where $\bm\varepsilon = \left[\varepsilon_1,\varepsilon_2,\cdots,\varepsilon_N\right]^{\sf T}$ is the dual variable vector associated with (\ref{eq:cvx_cond}).
The function $\mathcal G_1\left(\bm \theta, \bm\varepsilon \right)$ is convex and can be solved by the CVX solver in each iteration of (\ref{eq:up1}).
We can perform the projection operation to update $\bm\theta$, as given by
\begin{align}
\bm{\theta}^{\circ} &= \underset{ \bm\theta }{\arg\min} \ \mathcal G_1\left(\bm \theta, \bm\varepsilon \right) = \left[  {\theta}^{\circ}_1, {\theta}^{\circ}_2, \cdots, {\theta}^{\circ}_N \right]^{\sf T}, \label{eq:up_theta1} \\
\bm{\theta}^{(l+1)} &= \left[ e^{- \jmath \mathsf{arg} \left\{ {\theta}^{\circ}_1 \right\}}, e^{- \jmath \mathsf{arg} \left\{ {\theta}^{\circ}_2 \right\}}, \cdots, e^{- \jmath \mathsf{arg} \left\{ {\theta}^{\circ}_N \right\}} \right]^{\sf T}. \label{eq:up_theta2}
\end{align}

Given $\bm{\theta}^{(l+1)}$ in (\ref{eq:up_theta2}), we can take Newton's method to update $\bm q$, as given by
\begin{align}
\boldsymbol{q}^{(i+1)} = \boldsymbol{q}^{(i)} - \iota \left[ \nabla^2 \mathcal G_2\left(\bm q^{(i)} \right) \right]^{-1} \nabla \mathcal G_2\left(\bm q^{(i)} \right), \label{up_q}
\end{align}
where $\iota$ is a step size of Newton's method and $\mathcal G_2\left(\bm q \right)=\sum_{i=1}^{K} J_{B,k} (\bm q)+\frac{\rho}{2}\left\Vert\bm{\theta}-\bm{q}+\bm{r}\right\Vert^{2}$. In order to avoid inverting the Hessian matrix in (\ref{up_q}), we adopt the Quasi-Newton method and update the inverse Hessian matrix at each iteration. The proposed SIMin fraction transform-based ADMM method is described in Algorithm \ref{alg:admm}.

\begin{algorithm}[htbp]
\small
\caption{The proposed SIMin fraction transform based ADMM framework.}
\label{alg:admm}
\begin{algorithmic}[1]
\REQUIRE {Set feasible values of $\{ \bm \beta^{(0)}, \bm \theta^{(0)}, \bm q^{(0)}, \bm r^{(0)}\}$ and iteration index $t=0$.}
\REPEAT
\STATE {Set $t \leftarrow t+1$;}
\STATE {With given $\bm \theta^{(t-1)}$, update $\beta_k^{(t)}$ using (\ref{eq:up_beta});}
\STATE {Set iteration index $l = 0$;}
\REPEAT
\STATE {Set $l \leftarrow l+1$;}
\STATE {Update $\bm \theta^{(l)}$ according to (\ref{eq:up_theta1}) and (\ref{eq:up_theta2});}
\STATE {Update $\bm q^{(l)}$ according to the Quasi-Newton method;}
\STATE {Perform $\bm{r}^{(l)} = \bm{r}^{(l-1)}+\bm{\theta}^{(l)}-\bm{q}^{(l)}$;}
\UNTIL {The value of $\Vert\bm \theta^{(l)} - \bm q^{(l)} \Vert$ converges.}
\STATE {Obtain $\bm \theta^{(t)}=\bm \theta^{(l)}$;}
\UNTIL {The function in (P4) converges.}
\STATE {Output $\bm \theta$ and set $\mathbf \Theta = \mathsf{diag}\{\bm \theta\}$.}
\end{algorithmic}
\end{algorithm}

\subsection{Manifold Optimization Scheme}
Although Algorithm 1 is suitable for the \emph{weighted sum of inverse SINR minimization} problem, the relaxation of the unit modulus constraints in (\ref{eq:cvx_cond}) incurs performance loss. Additionally, the fraction transform requires solving the extra auxiliary variables.
In this section, we revisit the passive beamforming problem. The aim of the passive beamforming is to find the IRS reflection coefficients which reduce the transmit powers of the users while satisfying their latency requirements.
The passive beamforming is reformulated to make $\Gamma_k$ larger than the minimum
protection ratio $\tau_k$. Hence, the reflection beamforming subproblem is transformed to a new optimization problem.
Specifically, with (\ref{eq:cast}), constraint (\ref{eq:main_cond}) is rewritten as
\begin{align}
p_k \left\vert b_{k,k} + {\mathbf g}_{k,k}^{\sf H} \bm \theta \right\vert^2 \nonumber\\
\ge \widetilde{T}_k \Big( \sum_{j\neq k}^{K} p_j &  \left\vert b_{k,j} + \mathbf g_{k,j}^{\sf H} \bm \theta \right\vert^2 + \sigma_{\text u}^2 \Vert \mathbf f_k \Vert^2  \Big), \ \forall k.
\label{eq:theta_con}
\end{align}
By unfolding the squared terms in (\ref{eq:theta_con}), we define the ``latency residual'' of the $k$-th user as
\begin{align}
\alpha_k &= p_k \left\vert b_{k,k} + {\mathbf g}_{k,k}^{\sf H} \bm \theta \right\vert^2  -  \widetilde{T}_k \Big( \sum_{j\neq k}^{K} p_j  \left\vert b_{k,j} + \mathbf g_{k,j}^{\sf H} \bm \theta \right\vert^2 + \sigma_{\text u}^2 \Vert \mathbf f_k \Vert^2  \Big), \nonumber \\
&= \bm \theta^{\sf H} \Big( p_k {\mathbf g}_{k,k} {\mathbf g}_{k,k}^{\sf H} - \widetilde{T}_k \sum_{j\neq k}^{K} p_j \mathbf g_{k,j}\mathbf g_{k,j}^{\sf H} \Big) \bm \theta \nonumber\\
&+ 2\mathsf{Re}\Big\{ \Big(p_k b_{k,k}^{\ast}{\mathbf g}_{k,k}^{\sf H} -\widetilde{T}_k \sum_{j\neq k}^{K} p_j b_{k,j}^{\ast}\mathbf g_{k,j}^{\sf H} \Big)\bm \theta \Big\}  \nonumber \\
& + p_k b_{k,k}^2 - \widetilde{T}_k \sum_{j\neq k}^{K} \left( p_j b_{k,j}^2+\Vert \mathbf f_k \Vert^2 \right),
\label{eq:residual}
\end{align}
where $\bm \alpha=\left[ \alpha_1,\alpha_2,\cdots, \alpha_K\right]^{\sf T}$.
To fulfill the latency requirements, the passive beamforming subproblem (\ref{eq:optimal_theta}) is transformed to a \emph{latency residual maximization} problem, i.e., $\underset{\bm \theta, \bm \alpha}{\operatorname{max}} \sum_{k=1}^{K} \alpha_k$.
If and only if $\alpha_k, \forall k$ takes non-negative values to maximize $\sum_{k=1}^{K} \alpha_k$, then $\bm\theta$ can maximize the latency residual to reduce the latency.
A straightforward solution is to convert constraints (\ref{eq:theta_con}) and the unit modulus constraints into quadratic constraints. By suppressing the rank-one constraint, problem (\ref{eq:optimal_theta}) becomes SDR problem which is typically solved by eigen-decomposition.
However, the SDR problem suffers from a critical drawback that the number of optimization variables increases quadratically with the number of IRS elements. For this reason, we develop a CCMO method which can solve directly the \emph{latency residual maximization} problem.

In light of the special geometry of the constraint $\vert \theta_{n} \vert = 1$, we resort to Riemannian-Geometric optimization tools \cite{AbsMahSep2008}. The feasible region of the \emph{latency residual maximization} problem constitutes a complex circle manifold. The manifold representation can also provide a relatively concise form.
Riemannian manifold optimization breaks the confinement of the Euclidean space to generalize the gradient descent on a Riemannian manifold geometrically specified by the constraints. The original constrained optimization problem can be transformed into an unconstrained optimization problem and minimized by Riemannian gradient descent.

Referring to the latency residual representation (\ref{eq:residual}), the subproblem (\ref{eq:optimal_theta}) can be rewritten as
\begin{subequations}
\begin{align}
(\text{P5}): \quad \underset{ {\bm \theta}}{\operatorname{max}} \quad &  {f}_0( {\bm \theta}) =  {\bm \theta}^{\sf H} \mathbf U  {\bm \theta} + 2 \mathsf{Re} \left( \bm \theta^{\sf H} \mathbf v \right) + C \nonumber \\
\text{s.t.} \quad & \vert \theta_n \vert = 1,  \quad \ n=1,2,\cdots, N,
\end{align}
\end{subequations}
where
\begin{align}
\mathbf U &= \sum_{k=1}^{K} \Big( p_k \mathbf g_{k,k} \mathbf g_{k,k}^{\sf H} - \widetilde{T}_k \sum_{j\neq k}^{K} p_j \mathbf g_{k,j} \mathbf g_{k,j}^{\sf H} \Big); \\
\mathbf v &= \sum_{k=1}^{K} \Big( p_k b_{k,k} \mathbf g_{k,k} - \widetilde{T}_k \sum_{j\neq k}^{K} p_j b_{k,j} \mathbf g_{k,j} \Big); \\
C &= \sum_{k=1}^{K} \Big[ p_k \left\vert b_{k,k} \right\vert^2 - \widetilde{T}_k \Big( \sum_{j\neq k}^{K} p_j \left\vert {b}_{k,j} \right\vert^2 + \sigma_{\text u}^2 \Vert \mathbf f_k \Vert^2 \Big) \Big].
\end{align}
According to the notion of manifold optimization, problem (P5) can be reformulated as:
\begin{align}
(\text{P6}): \quad \underset{ {\bm \theta} \in \mathcal S^{N}}{\operatorname{min}} \quad & f( {\bm \theta}) = -{\bm \theta}^{\sf H} \mathbf U {\bm \theta} - 2 \mathsf{Re} \left( \bm \theta^{\sf H} \mathbf v \right),
\end{align}
where $\mathcal S^{N}$ denotes the manifold space defined by the constant modulus constraints, i.e.,
\begin{align}
\mathcal S^{N} = \left\{ {\bm \theta} \in \mathbbmss{C}^{N}: \vert \theta_1 \vert = \vert \theta_2 \vert = \cdots = \vert \theta_{N} \vert = 1  \right\},
\end{align}
where $\mathcal S = \left\{ \theta_n \in \mathbbmss{C}: \theta_n \theta_n^{\ast} = \mathsf{Re}\{\theta_n\}^2+\mathsf{Im}\{\theta_n\}^2 = 1  \right\}$ is known as a complex circle and can be viewed as a sub-manifold of $\mathbbmss C$. The search space $\mathcal S^{N}$ is the product of $N$ complex circles. Referred to as a complex circular manifold, the search space $\mathcal{S}^N$ is a sub-manifold of $\mathbbmss C^{N}$.

\begin{figure}[t]
	\centering{}\includegraphics[scale=0.35]{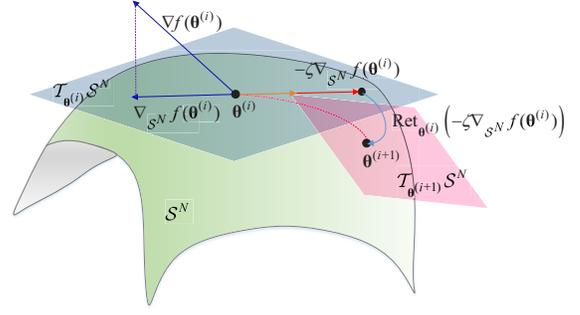}
	\caption{An illustration of gradient descent on Riemannian manifolds.}
	\label{fig:manopt}
\end{figure}

The new CCMO algorithm performs gradient descent on the complex circular manifold. The gradient descent on the Riemannian manifold is similar to that in the Euclidean space. It consists of two phases: finding the descent direction of the current solution by computing the negative Riemannian gradient, and decreasing the value of the objective function via \emph{line search} \cite{AbsMahSep2008}. The Riemannian gradient of $f({\bm \theta})$ at the current iteration point ${\bm \theta}^{(i)} \in \mathcal S^{N}$ is the projection of the search direction in the Euclidean space onto the tangent space $\mathcal T_{{\bm \theta}^{(i)}} \mathcal S^{N}$, as given by
\begin{align}
\mathcal T_{{\bm \theta}^{(i)}} \mathcal S^{N} = \left\{ \bm \eta \in \mathbbmss{C}^{N+1}: \mathsf{Re} \{ \bm \eta^{\ast} \odot {\bm \theta}^{(i)} \} = \mathbf 0  \right\}.
\end{align}
The Euclidean gradient of $f({\bm \theta}^{(i)})$ at ${\bm \theta}^{(i)}$ is given by
\begin{align}
\nabla f({\bm \theta}^{(i)}) = - 2 \mathbf U {\bm \theta^{(i)}} - 2 \mathbf v.  \label{eq:eucl_gradient}
\end{align}
By performing a \emph{projection operator}, the Riemannian gradient of $f({\bm \theta}^{(i)})$ is obtained as
\begin{align}
\nabla_{\mathcal S^{N}} f({\bm \theta}^{(i)}) &= \mathsf{Proj}_{\mathcal T_{{\bm \theta}^{(i)}} \mathcal S^{N} } \left( \nabla f({\bm \theta}^{(i)}) \right)
\nonumber\\
&= \nabla f({\bm \theta}^{(i)}) - \mathsf{Re} \{ \nabla f({\bm \theta}^{(i)})^{\ast} \odot {\bm \theta}^{(i)} \} \odot {\bm \theta}^{(i)}. \label{eq:riem_gradient}
\end{align}
Hence, the current point ${\bm \theta}^{(i)}$ in the tangent space $\mathcal T_{{\bm \theta}^{(i)}} \mathcal S^{N}$ is updated as
\begin{align}
{{\bm \theta}^{(i)}}^{\circ} = {\bm \theta}^{(i)} - \zeta \nabla_{\mathcal S^{N}} f({\bm \theta}^{(i)}), \label{eq:gd_over}
\end{align}
where $\zeta > 0$ is a carefully selected constant step size\footnote{To ensure stability and convergence of the CCMO algorithm, the step size $\zeta$ should be selected to satisfy $\zeta \le 1/\lambda_{\mathbf U}$ where $\lambda_{\mathbf U}$ represents the largest eigenvalue of the matrix $\mathbf U$ in problem (P5). This optimization problem can be solved by leveraging the Manopt toolbox in MATLAB \cite{8125771, 8706630, manopt}.}. We note that ${{\bm \theta}^{(i)}}^{\circ}$ is still in the tangent space $\mathcal T_{{\bm \theta}^{(i)}} \mathcal S^{N+1}$ but may not be on the manifold $\mathcal S^{N}$. Therefore, a \emph{Retraction mapping} operation is applied to move the point ${{\bm \theta}^{(i)}}^{\circ}$ back onto the manifold $\mathcal S^{N}$. Finally, the point ${\bm \theta}^{(i+1)}$ is updated by using the \emph{Retraction mapping}, as given by
\begin{align}
{\bm \theta}^{(i+1)} &= \mathsf{Ret}_{{{\bm \theta}^{(i)}}} \left( - \zeta \nabla_{\mathcal S^{N}} f({\bm \theta}^{(i)}) \right) \nonumber\\
&= \frac{{\bm \theta}^{(i)} - \zeta \nabla_{\mathcal S^{N}} f({\bm \theta}^{(i)})}{\Vert {\bm \theta}^{(i)} - \zeta \nabla_{\mathcal S^{N}} f({\bm \theta}^{(i)}) \Vert}
= {{\bm \theta}^{(i)}}^{\circ} \odot \frac{1}{\vert {{\bm \theta}^{(i)}}^{\circ} \vert}. \label{eq:retrac}
\end{align}
The above operations are illustrated in Fig. \ref{fig:manopt} and summarized in Algorithm \ref{alg:ccm}.

\begin{algorithm}[htbp]
\small
\caption{The proposed CCMO algorithm for passive beamforming.}
\label{alg:ccm}
\begin{algorithmic}[1]
\REQUIRE {Set feasible values of $\{ \bm \theta^{(0)}\}$ and iteration index $i=0$.}
\REPEAT
\STATE {Set $i \leftarrow i+1$;}
\STATE {Calculate the Euclidean gradient $\nabla f({\bm \theta}^{(i)})$ at ${\bm \theta}^{(i)}$ using (\ref{eq:eucl_gradient});}
\STATE {Construct the tangent space $\mathcal T_{{\bm \theta}^{(i)}} \mathcal S^{N}$ and calculate the current Riemannian gradient $\nabla_{\mathcal S^{N}} f({\bm \theta}^{(i)})$ using (\ref{eq:riem_gradient});}
\STATE {Perform gradient descent algorithm over the current tangent space using (\ref{eq:gd_over});}
\STATE {Update ${\bm \theta}^{(i+1)}$ using the Retraction mapping operator according to (\ref{eq:retrac});}
\UNTIL {The value of $\vert f({\bm \theta}^{(i)}) -f( {\bm \theta}^{(i-1)}) \vert$ in (P6) converges.}
\STATE {Output $\bm \theta$ and obtain $\mathbf \Theta = \mathsf{diag}\{\bm \theta\}$.}
\end{algorithmic}
\end{algorithm}

\subsection{Complexity Analysis}

In terms of computational complexity, the key difference between the two proposed methods and the aforementioned SDR-based baseline is in passive beamforming. At each iteration of passive beamforming, the SDR-based baseline described in Section IV-B incurs a high complexity of $\mathcal{O}(N^{6.5})$ \cite{8706630}, since it increases the number of variables. For the proposed SIMin fraction transform-based ADMM described in Algorithm 2, the complexity is dominated by (35) and (36). Problem (35) is a convex quartic function and solved by CVX. The complexity of solving (35) is $\mathcal{O}(N^{3.5})$ \cite{9039554}. The Quasi-Newton method involved in (36) adopts the Broyden-Fletcher-Goldfarb-Shanno (BFGS) method \cite{broyden1970}, which has a complexity of $\mathcal{O}(N^{2})$. As a result, the complexity of Algorithm 2 is $\mathcal{O}(N^{3.5})$ per iteration. For the proposed CCMO algorithm described in Algorithm 3, the complexity is dominated by the evaluation of the Riemannian gradient descent which is $\mathcal{O}(N^{2})$ per iteration \cite{8706630}. As a result, the complexity of Algorithm 3 is $\mathcal{O}(N^2)$ per iteration. To this end, the two proposed algorithms are orders of magnitude lower than the SDR-based alternative.

\subsection{Extension to multi-antenna users}

Consider a mmWave system with multi-antenna users, each user equipped with $N_u$ antennas. The channel between the AP and the $k$-th user is $\mathbf{H}_{\text{d},k}\in \mathbbmss{C}^{M \times N_u}$ and the channel between the IRS and the $k$-th user is $\mathbf{H}_{\text{r},k}\in \mathbbmss{C}^{N \times N_u}$. Then, the received signal at the AP from the k-th user can be rewritten as
\begin{align}
y_k =\mathbf f_k^{\sf H} \Big( \sum_{j= 1}^{K} \left(\mathbf H_{\text{d},j} + \mathbf G \mathbf{\Theta} \mathbf H_{\text{r},j} \right) \mathbf{q}_j  {s}_j + \mathbf{u}_k \Big),
\label{eq:signal_mu}
\end{align}
where $\mathbf{q}_j$ is the uplink transmit beamformer of the $j$-th user and $\mathbf{Q}=[\mathbf{q}_1, \cdots, \mathbf{q}_K] \in \mathbbmss{C}^{N_u \times K}$. The SINR of the $k$-th user becomes
\begin{equation}
\Gamma_k(\mathbf Q, \mathbf F, \mathbf \Theta) = \frac{ \left\vert \mathbf f_k^{\sf H} \left(\mathbf H_{\text{d},k} + \mathbf G \mathbf{\Theta} \mathbf H_{\text{r},k} \right) \mathbf{q}_k \right\vert^2}{ \sum_{j\neq k}^{K}  \left\vert \mathbf f_k^{\sf H} \left(\mathbf H_{\text{d},j} + \mathbf G \mathbf{\Theta} \mathbf H_{\text{r},j}\right) \mathbf{q}_j \right\vert^2 + \sigma_{\text u}^2 \Vert \mathbf f_k \Vert^2 }.
\label{eq:sinr_mu}
\end{equation}
As a result, the problem of interest is written as
\begin{align}
(\text{P7}): \quad & \underset{\mathbf{Q}, \mathbf{F}, \mathbf{\Theta}}{\operatorname{min}} \quad  \sum_{k=1}^{K} \Vert \mathbf{q}_k \Vert^2 \nonumber  \\
\text{s.t.}  & \quad \frac{D_k}{ W \log \left(1 + \Gamma_k(\mathbf Q, \mathbf F, \mathbf \Theta) \right)} \le T, \ \forall k,\label{eq:MIN3_mu} \\
\quad &   \quad \text{(\ref{eq:MIN2})}. \nonumber
\end{align}
Problem (P7) is a joint optimization of the transmit beamformers at the users, the multi-user detectors at the AP, and the phase shifts at the IRS. With the proposed alternating optimization framework, the IRS configuration and multi-user detectors can be solved by the proposed algorithms. Specifically, given the IRS configuration $\mathbf{\Theta}$ and the transmit beamformer $\mathbf{q}_k$, the effective channel $\mathbf{h}_k=\left(\mathbf H_{\text{d},k} + \mathbf G \mathbf{\Theta} \mathbf H_{\text{r},k} \right) \mathbf{q}_k$ is given, where $\mathbf h_{\text{d},k}=\mathbf H_{\text{d},k}\mathbf{q}_k$ and $\mathbf h_{\text{r},k}=\mathbf H_{\text{r},k}\mathbf{q}_k$. The optimization of the multi-user detectors can still be solved by the MVDR method, as done in (\ref{eq:update_f}). Likewise, when the transmit beamformer $\mathbf{q}_k$ and multi-user detectors $\mathbf{F}$ are given, we have $\mathbf{\Theta} \mathbf H_{\text{r},k}\mathbf{q}_k = \mathsf{diag}(\mathbf h_{\text{r},k}) \bm\theta$, and the proposed passive beamforming algorithms, i.e., Algorithms 2 and 3, remain valid. When $\mathbf{F}$ and $\mathbf{\Theta}$ are given, the effective channel becomes $\mathbf{w}_{k,j}=\left(\mathbf H_{\text{d},k} + \mathbf G \mathbf{\Theta} \mathbf H_{\text{r},k} \right) ^{\sf H} \mathbf f_k$ for notation brevity. The normalized uplink beamformer at the $k$-th user is denoted by $\bar{\mathbf q}_k = \frac{{\mathbf q}_k}{\Vert {\mathbf q}_k\Vert }$. Problem (P7) can be recast to optimizing $\mathbf{q}_k$, as given by
\begin{subequations}
\begin{align}
&(\text{P8}): \quad  \underset{\bar{\mathbf q}_k, p_k}{\operatorname{min}} \quad  \sum_{k=1}^{K} p_k \nonumber  \\
&\text{s.t.}  \quad    \Vert \bar{\mathbf q}_k \Vert^2 =1,  \label{eq:norm_q}
\end{align}
\end{subequations}
\begin{align}
-p_k &\left\vert \bar{\mathbf q}_k^{\mathsf H}  \mathbf w_{k,k}  \right\vert^2 + \widetilde{T}_k \Big( \sum_{j\neq k}^{K} p_j  \left\vert \bar{\mathbf q}_j^{\mathsf H} \mathbf w_{k,j}  \right\vert^2 \nonumber \\
& \quad\quad+ \sigma_{\text u}^2 \Vert \mathbf f_k \Vert^2 \Big) \le  0, \ \forall k, \tag{59b} \label{eq:cond_mu}
\end{align}
where $p_k=\Vert {\mathbf q}_k\Vert^2$ is the transmit power of the $k$-th user, and (\ref{eq:cond_mu}) can be referred to (\ref{eq:main_cond}). Problem (P8) displays strong analogy to problem (P1), and can be transformed to a Rayleigh quotient in the same way as problem (P3). The Rayleigh quotient is invariant to the scaling of $\bar{\mathbf q}_k$. Therefore, problem (P8) can be solved by Algorithm 1, followed by a normalization of the resultant $\bar{\mathbf{q}}_k$, as required in (\ref{eq:norm_q}).

\section{Numerical And Simulation Results}\label{sec:sim}
In this section, simulations are carried out to verify the proposed uplink transmit power control methods and potential benefits of deploying IRS in mmWave SIMO systems.
\subsection{Simulation Setup}
\begin{figure}[t]
	\centering{}\includegraphics[scale=0.5]{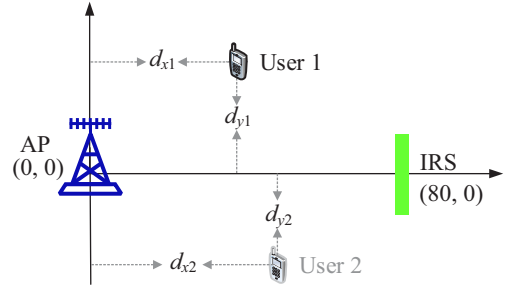}
	\caption{The simulated IRS-assisted mmWave SIMO scenario.}
	\label{fig:layout}
\end{figure}

We consider the IRS-aided mmWave system in Fig. \ref{fig:layout}, where the AP is equipped with a ULA of $M=32$ antennas, and located at the $(x,y)$-coordinates of $(0,0)$. The IRS is implemented with a URA, where the vertical length is set to $N_{\rm az}=5$ and the horizontal length $N_{\rm el}$ varies in different simulations. The center of the IRS is at $(80\ \text{m}, 0)$.
Both single-user and multi-user scenarios are investigated. In the single-user scenario, we assume that there is only User 1 in the system and its position is $(d_{x1}, d_{y1})$. In the multi-user scenario, both User 1 and User 2 transmit, and the position of User 2 is $(d_{x2}, -d_{y2})$.

According to channel measurements \cite{6834753, wang2019intelligent}, the channel gain $\xi$ follows a complex Gaussian distribution:
\begin{align}
\xi \sim \mathcal{CN} ( 0, 10^{- \text{PL} \left( R \right) }),
\end{align}
where $\text{PL} \left( R \right)$ is the path-loss (in dB) over distance $R$ (in meters), as given by
\begin{align}
\text{PL} \left( R \right) = \chi_a + 10 \chi_b \log_{10} \left( R \right) + \kappa,
\end{align}
and $\kappa \sim \mathcal{N}\left(0, \sigma_{\kappa}^2 \right)$ accounts for the lognormal shadowing.
For mmWave communications at $28\ \text{GHz}$, the channel gain is generated in two cases. In the case of LoS path, the parameter values are $\chi_a = 61.4$, $\chi_b = 2$ and $\sigma_{\kappa}=5.8\ \text{dB}$. In the case of NLoS path, $\chi_a = 72$, $\chi_b = 2.92$ and $\sigma_{\kappa}=8.7\ \text{dB}$. To investigate the role of the IRS in the uplink transmission, the mmWave channels between the AP and users are categorized into the following two scenarios:
\begin{itemize}
\item LoS scenario, where only LoS signal is received.
\item Obstructed-line-of-sight (OLoS) scenario, where the LoS component is blocked and only NLoS components exist. We use $\rho_{\text b}$ to indicate the blockage probability of the LoS path, or in other words, the OLoS probability.
\end{itemize}
According to \cite{wang2019intelligent} and \cite{8207426}, in typical uplink mmWave systems, the antenna gains of the user and AP are set to $\varrho_{\text U}=0\ \text{dBi}$ and $\varrho_{\text B}=9.82\ \text{dBi}$, respectively.
%We take $\varrho_{\text U}=0\ \text{dBi}$ for IRS-user link, and $\varrho_{\text B}=9.82\ \text{dBi}$ for IRS-AP link.
Generally, the IRS-reflected channel is relatively weak due to the \emph{double-fading} effect \cite{5162013} and high path loss of mmWave. The reflection gain of the IRS elements can compensate for channel attenuation, and therefore, the relative reflection gain of the IRS is $\nu=\frac{\varrho_{\text I}}{\sqrt{\varrho_{\text B} \varrho_{\text U}}}$ \cite{2019arXiv190507920G}.

Unless otherwise specified, the rest of the simulation parameters are as follows: the noise variance is $\sigma_{\text u}^2=-85\ \text{dBm}$; the system bandwidth is $W=500\ \text{MHz}$; the latency requirement is $T=50 \ \text{ms}$; the relative reflection gain is $\nu=15\ \text{dB}$; the location parameters are $d_{x1}=40\ \text{m}$, $d_{y1}=40\ \text{m}$, $d_{x2}=50\ \text{m}$, and $d_{y2}=20\ \text{m}$; the number of NLoS paths is $L=3$; and the data of each user follows the uniform distribution $D_k \sim \mathcal{U}\left( 5000\ \text{nats}, 8000\ \text{nats}\right)$.

\subsection{Single-User Scenario}

For comparison, two baseline schemes are considered:
\begin{itemize}
\item \textbf{Baseline without IRS}: We consider the uplink transmit power allocation in the absence of IRS. This baseline is a reduced version of the proposed Algorithm 1, with no passive beamforming required.
\item \textbf{SDR}: The SDR is adopted to obtain the reflection coefficients of the IRS, where the rank-one approximation \cite{5447068} is applied to obtain a feasible solution, as described in Section IV-B.
\end{itemize}

\begin{figure}[t]
    \centering
    \subfigure[]{\includegraphics[width=2.6in]{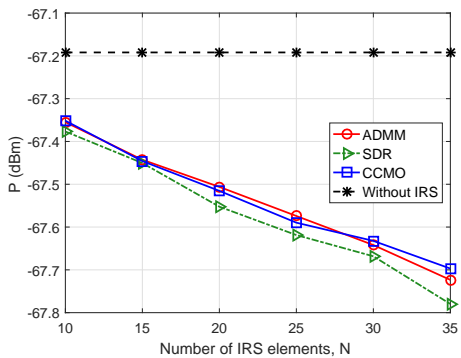}}
    \quad
    \subfigure[]{\includegraphics[width=2.6in]{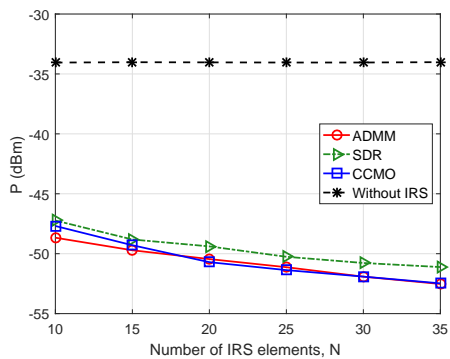}}
    \caption{Transmit power versus number of IRS elements, $N$. (a) LoS scenario; (b) OLoS scenario.}
    \label{fig:1_n}
\end{figure}

First, we investigate the impact of the number of IRS elements on the uplink transmit power.
Figs. \ref{fig:1_n}(a) and 4(b) show the allocated transmit power of the user in the LoS and OLoS scenarios, respectively. As expected, the power allocated in the absence of IRS is consistent in both scenarios. All the three schemes with the IRS outperform the baseline without IRS.
It is observed that the schemes with the IRS can substantially reduce the transmit power in the OLoS scenario, whereas only a slight improvement can be achieved in the LoS scenario. This is because the direct LoS link offers a considerably higher channel gain than the supplementary IRS-reflected link via the IRS, due to double-fading effect and high path loss of the IRS-reflected channel. In the presence of the dominant LoS path, the three schemes with the IRS provide the same performance.
However, Fig. \ref{fig:1_n}(b) reveals that the supplementary IRS-reflected link is dominant in the OLoS scenario, thanks to the IRS-enhanced received power.
It is also seen that the optimal user power decreases with the increase of $N$, as the result of the increasing passive beamforming gain. The proposed CCMO and ADMM methods stably outperform the SDR-based alternative, especially in the OLoS scenarios.

\begin{figure}[t]
	\centering
    \subfigure[]{\includegraphics[width=2.6in]{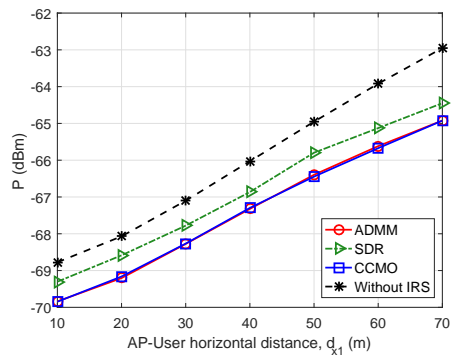}}
    \quad
    \subfigure[]{\includegraphics[width=2.6in]{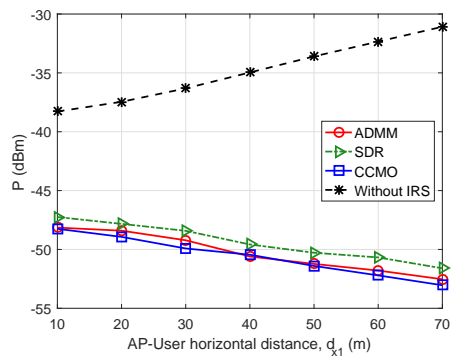}}
	\caption{Transmit power versus AP-user horizontal distance, $d_{x1}$. (a) LoS scenario; (b) OLoS scenario.}
	\label{fig:1_dis}
\end{figure}

Fig. \ref{fig:1_dis} examines the uplink transmit power versus the horizontal distance between the AP and User 1. As $d_{x1}$ increases from 10 to 70, the distance between the AP and User 1 increases and the distance between the IRS and User 1 decreases. The power required by the scheme without IRS increases rapidly, as the user moves away from the AP.
The gains of the proposed ADMM and CCMO algorithms over the two baselines are shown in Fig. \ref{fig:1_dis}. In Fig. \ref{fig:1_dis}(a), the performance gaps between the scheme without IRS and the IRS-aided schemes grow increasingly with the distance between the AP and user. This is because the gain of the IRS becomes increasingly prominent, as the propagation loss between the AP and user grows. In Fig. \ref{fig:1_dis}(b), the performance gaps enlarge for the same reason.

\begin{figure}[t]
	\centering
    \subfigure[]{\includegraphics[width=2.6in]{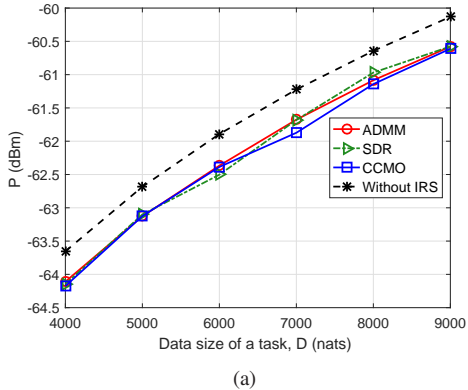}}
    \quad
    \subfigure[]{\includegraphics[width=2.6in]{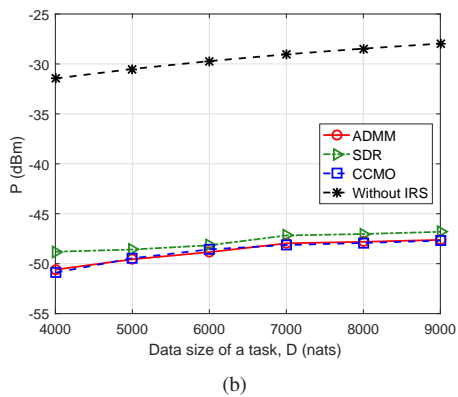}}
	\caption{Transmit power versus the data size, $D$. (a) LoS scenario; (b) OLoS scenario.}
	\label{fig:1_data}
\end{figure}

Fig. \ref{fig:1_data}(a) shows the powers achieved by all the schemes increase, as the amount of data enlarges in the LOS scenario. As expected, the power required by the IRS-aided system is lower than the system without IRS. The power gap enlarges in the OLoS scenario, as shown in Fig. \ref{fig:1_data}(b). We also observe that the proposed CCMO and ADMM methods significantly outperform the SDR method, which is consistent with Fig. \ref{fig:1_n}.

\begin{figure}[t]
	\centering{}\includegraphics[width=2.6in]{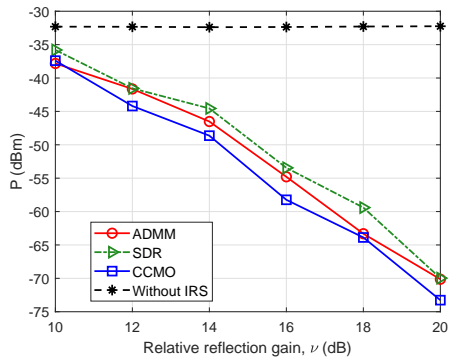}
	\caption{Transmit power versus the relative reflection gain, $\nu$.}
	\label{fig:1_nlos_gain}
\end{figure}

In Fig. \ref{fig:1_nlos_gain}, we assess the impact of the relative reflection gain of the IRS on the transmit power of the user. When the relative reflection gain grows from 10 to 20 dB, all the considered schemes with IRS can reduce considerably the transmit power of the user. The reason is that the channel quality of the IRS-aided links improves with the increasing reflection gain.

\begin{figure}[t]
	\centering{}\includegraphics[width=2.6in]{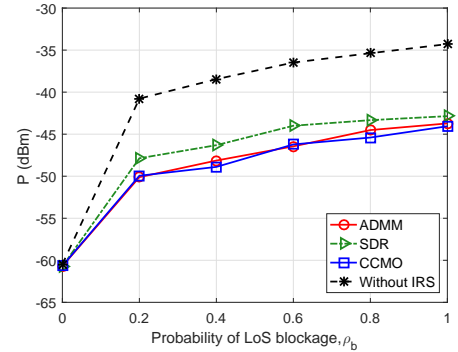}
	\caption{Transmit power versus the probability of LoS blockage, $\rho_{\text b}$.}
	\label{fig:1_nlos_prob}
\end{figure}

Fig. \ref{fig:1_nlos_prob} plots the achievable minimum user power under different LoS blockage probabilities. It shows that a larger blockage probability results in higher transmit powers under all schemes. In the case of $\rho_{\text b}=0$ indicating the LoS scenario, the performance gain attributed to the IRS is negligible. Conversely, in the case of $\rho_{\text b}=1$ indicating the OLoS scenario, the proposed IRS-assisted uplink mmWave system design saves the transmit power while meeting the latency requirements. Fig. \ref{fig:1_nlos_prob} also confirms the improvement of the proposed ADMM and CCMO algorithms over the SDR-based alternative methods.

\subsection{Multi-User Scenario}
We consider two simultaneous users, i.e., User 1 and User 2. We assume that the blockages of the two users are statistically independent, as modeled in \cite{5733382}.

\begin{figure}[t]
	\centering{}\includegraphics[width=3.2in]{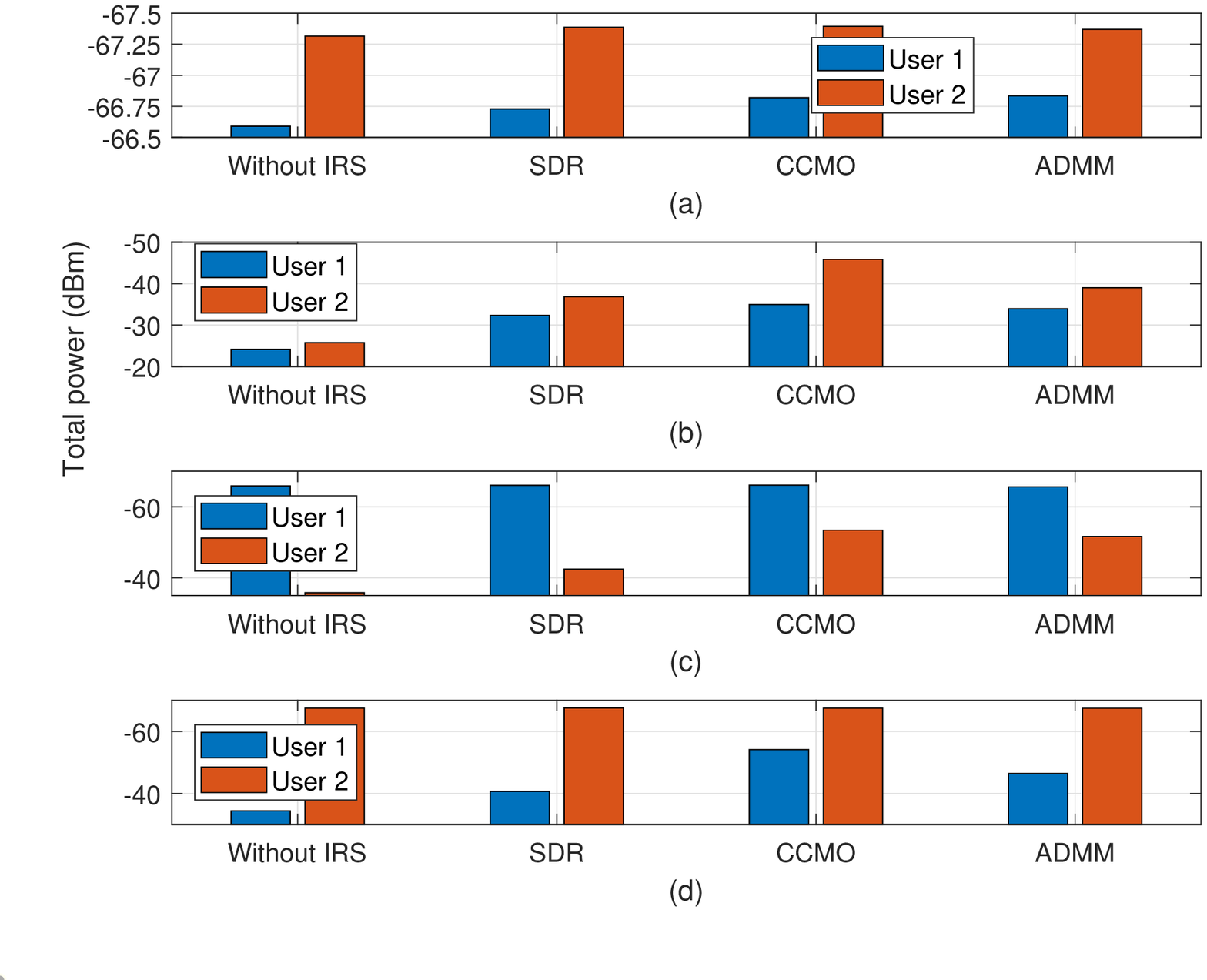}
	\caption{Comparison of transmit power between the two users. (a) Both users have LoS. (b) Both users have OLoS. (c) User 1 has LoS and User 2 has OLoS; (d) User 1 has OLoS and User 2 has LoS.}
	\label{fig:two_user}
\end{figure}

Fig. \ref{fig:two_user}(a) considers that both two users have LoS links. The schemes with the IRS provide a little improvement, as compared with the baseline without IRS. Also, User 1 consumes more power than User 2, because $\widetilde{T}_1 \ge \widetilde{T}_2$. Fig. \ref{fig:two_user}(b) considers that both two users undergo blockages, and the IRS-reflected link plays a key role to reduce the transmit powers of the users.

In Fig. \ref{fig:two_user}(c), we consider that User 1 has a LoS link while User 2 is blocked. We see a substantial gain of the schemes with the IRS for User 2, whereas the IRS has little effect on the performance of User 1. This validates that the benefit of the IRS is increasingly prominent, as the direct links become weak. Although $\widetilde{T}_1$ is higher than $\widetilde{T}_2$ (i.e., User 1 has a higher service priority), User 1 has a lower transmit power than User 2 due to the LoS availability of User 1.
Fig. \ref{fig:two_user}(d) shows the case where User 1 is blocked while User 2 enjoys the LoS link. The same observation can be made as in Fig. \ref{fig:two_user}(c), by switching the roles of Users 1 and 2.

\begin{figure}[t]
	\centering{}\includegraphics[width=2.6in]{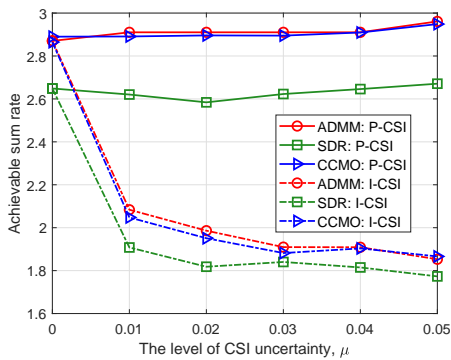}
	\caption{Comparison of sum rate between perfect and imperfect CSI.}
	\label{fig:csi}
\end{figure}

\begin{figure}[t]
	\centering{}\includegraphics[width=2.6in]{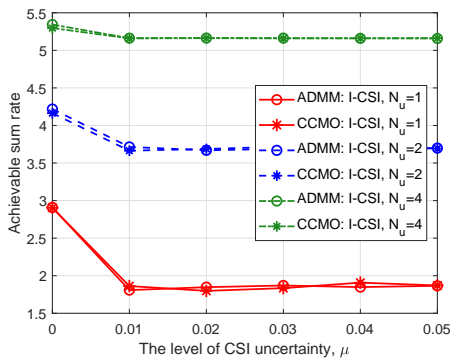}
	\caption{The sum rate versus $N_u$ under imperfect CSI.}
	\label{fig:multi_ant}
\end{figure}

\begin{figure}[t]
	\centering
    \subfigure[]{\includegraphics[width=2.6in]{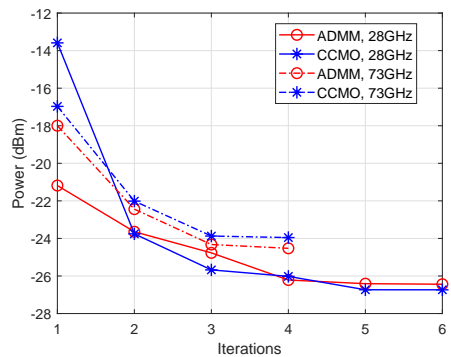}}
    \quad
    \subfigure[]{\includegraphics[width=2.6in]{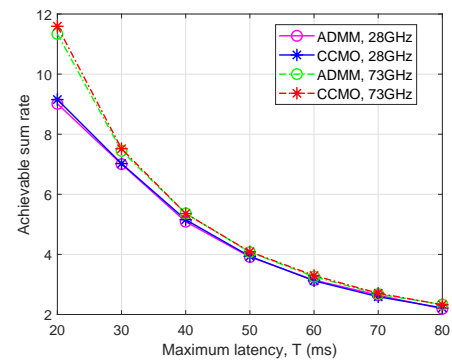}}
	\caption{The algorithm performances with different frequencies. (a) Convergence behavior; (b) Achievable sum rate.}
	\label{fig:freq}
\end{figure}

Fig. \ref{fig:csi} evaluates the sum rate of the considered algorithms under both imperfect CSI (I-CSI) and perfect CSI (P-CSI). Since the reflection channel from the IRS to the user is more difficult to estimate than the direct channel from the AP to the user and the indirect channel from the AP to the IRS \cite{9110587, 9217312}, we assume that the reflection channel $\mathbf{h}_{\text{r},k}$ is imperfect and modeled as $\mathbf{h}_{\text{r},k}=\hat{\mathbf{h}}_{\text{r},k} + \Delta \mathbf{h}_k$, where $\Delta \mathbf{h}_k$ is the CSI error vector and $\hat{\mathbf{h}}_{\text{r},k}$ is the estimate of $\mathbf{h}_{\text{r},k}$. Based on the channel error bounded model \cite{9110587}, we define $\Vert \Delta \mathbf{h}_k \Vert_2 \le \mu \Vert \hat{\mathbf{h}}_{\text{r},k}\Vert_2$ where $\mu$ is the level of CSI uncertainty.
As shown in Fig. \ref{fig:csi}, the uplink sum rate degrades under I-CSI, as compared to P-CSI. Nevertheless, the proposed algorithms are shown to still outperform the SDR-based alternative, especially when the CSI uncertainty level is low and the channels are reasonably well estimated.
Fig. \ref{fig:multi_ant} shows the impact of the number of user antennas on the sum rate under I-CSI. We see that the actual achievable sum rate grows with the number of user antennas, $N_u$.

Fig. \ref{fig:freq}(a) plots the convergence behaviors of the proposed methods when the operating frequency of the system is 28 and 73 GHz. We see that both of the two proposed algorithms behave similarly. The required total power is higher at 73 GHz than it is at 28 GHz. This is because higher frequencies incur larger path losses. Furthermore, the proposed algorithms converge faster at 73 GHz than they do at 28 GHz.
Fig. \ref{fig:freq}(b) investigates the achievable sum rate under different latency requirements. We observe that our algorithms provide the same sum-rate performances at both 28 GHz and 73 GHz, under the latency requirement of more than 40 ms. The achievable sum rate is higher at 73 GHz than it is at 28 GHz, especially under the latency requirement of 20 ms.

\section{Conclusion}\label{sec:con}
In this paper, we minimized the transmit power of the users while satisfying their latency requirements, by jointly optimizing the transmit powers, the multi-user detector at the AP, and the passive beamforming coefficients at the IRS. We designed an alternating optimization framework to transform the joint optimization problem into three tractable subproblems. Closed-form solutions were derived for the transmit powers and multi-user detector. We also developed two algorithms to efficiently deliver the passive beamforming coefficients of IRS. Simulations validated the benefits of deploying the IRS in the uplink mmWave SIMO system. Insights were shed on the role of the IRS in the uplink power allocation.

\begin{appendices}
\section{Proof of Lemma 1}\label{prf1}

Assume that two sets of positive power vectors $\hat{\mathbf p}$ and $\mathbf p^{\star}$ are the fixed points of $\mathcal M_{\mathbf{f}_k}(\mathbf{p})$. Without loss of generality, we suppose that there exists an index $k$ satisfying $\hat{p}_k > p^{\star}_k$, and let $\max_{j} \left(\hat{p}_j / p^{\star}_j \right) =\gamma > 1$. Thus $\gamma \mathbf p^{\star} \ge \hat{\mathbf p}$.
We can find such an index $i$ that $\gamma p^{\star}_i = \hat{p}_i$. Since both $\hat{\mathbf p}$ and $\mathbf p^{\star}$ are the fixed points of $\mathcal M_{\mathbf{f}_k}$, we have
\begin{align}
\hat{p}_i &= \underset{\mathbf f_i}{\operatorname{min}} \Big\{\sum_{j\neq i}^{K} \frac{ \widetilde{T}_i \vert \mathbf f_i^{\sf H} \mathbf h_j \vert^2}{\vert \mathbf f_i^{\sf H} \mathbf h_i  \vert^2} \hat{p}_j  + \frac{\sigma_{\text u}^2 \widetilde{T}_i \Vert \mathbf f_i \Vert^2 }{ \vert \mathbf f_i^{\sf H} \mathbf h_i  \vert^2}\Big\}  \nonumber\\
& \le \underset{\mathbf f_i}{\operatorname{min}} \Big\{\sum_{j\neq i}^{K} \frac{ \widetilde{T}_i \vert \mathbf f_i^{\sf H} \mathbf h_j \vert^2}{\vert \mathbf f_i^{\sf H} \mathbf h_i  \vert^2} \gamma p^{\star}_j  + \frac{\sigma_{\text u}^2 \widetilde{T}_i \Vert \mathbf f_i \Vert^2}{ \vert \mathbf f_i^{\sf H} \mathbf h_i  \vert^2}\Big\}  \nonumber\\
& < \gamma \Big(\underset{\mathbf f_i}{\operatorname{min}} \Big\{\sum_{j\neq i}^{K} \frac{ \widetilde{T}_i \vert \mathbf f_i^{\sf H} \mathbf h_j \vert^2}{\vert \mathbf f_i^{\sf H} \mathbf h_i  \vert^2} p^{\star}_j  + \frac{\sigma_{\text u}^2 \widetilde{T}_i \Vert \mathbf f_i \Vert^2}{ \vert \mathbf f_i^{\sf H} \mathbf h_i  \vert^2}\Big\} \Big)  \nonumber\\
&= \gamma p^{\star}_i ,
\end{align}
which contradicts with $\gamma p^{\star}_i=\hat{p}_i$.
We conclude that the fixed point of $\mathcal M_{\mathbf{f}_k}(\mathbf{p})$ is unique. After the BCD iterations, we have $\mathcal{P}(\mathbf{F}^{(t-1)}, \bm{\theta}^{(t-1)}) \ge\mathcal{P}(\mathbf{F}^{(t-1)}, \bm{\theta}^{t}) \ge \mathcal{P}(\mathbf{F}^{t}, \bm{\theta}^{t})$. Furthermore, the uplink power is lower bounded due to the stringent latency  requirements. \QEDA%\qed

\section{Proof of Theorem 1}\label{prf2}
Let us introduce the following equivalence relationship:
\begin{align}
&z_k {A_k(\bm \theta)}^2 + \frac{1}{4 z_k} \frac{1}{B_k(\bm \theta)^2} \nonumber\\
=&\left( \sqrt{z_k} {A_k(\bm \theta)} - \frac{1}{2 \sqrt{z_k} } \frac{1}{B_k(\bm \theta)} \right)^2 + \frac{A_k(\bm \theta)}{B_k(\bm \theta)}. \label{eq:min_fp}
\end{align}
It is facile to see that minimizing the right-hand side (RHS) of (\ref{eq:min_fp}) with respect to $\bm z$ and $\bm \theta$ is equivalent to the minimization of the left-hand side (LHS). Meanwhile, we observe that the optimal $\bm z$ minimizing the RHS of (\ref{eq:min_fp}) can be found by
\begin{align}
z_k = \frac{1}{ 2 A_k(\bm \theta) B_k(\bm \theta)},
\end{align}
which forces the squared term in (\ref{eq:min_fp}) to be zero. Therefore, the optimal $\bm \theta$ for the LHS of (\ref{eq:min_fp}) is part of the solutions for minimizing $\frac{A_k(\bm \theta)}{B_k(\bm \theta)}$. \QEDA%\qed
\end{appendices}

\bibliographystyle{IEEEtran}
\bibliography{ciations}

\end{document}